\documentclass[]{JHEP3} % 10pt is ignored!
\usepackage{epsfig,multicol}

\def\g0{\gamma_0}
\def\gp{\gamma^{+}}
\def\gm{\gamma^{-}}

\def\nn{\nonumber}
\def\beqa{\begin{eqnarray}}
\def\eeqa{\end{eqnarray}}
\def\beqn{\begin{eqnarray}}
\def\eeqn{\end{eqnarray}}
\def\beq{\begin{equation}}
\def\eeq{\end{equation}}

\def\e{\epsilon}
\def\cG{r_\Gamma}
\def\I4{I_4}
\def\pu{p_1}
\def\pd{p_2}
\def\pt{p_3}
\def\pq{p_4}
\def\li{{\rm Li_2}}
\def\Li{{\rm Li_2}}

\def\Ll{{\rm L}}
\def\sud{s_{12}}
\def\sdt{s_{23}}

\def\rr{\right) }
\def\lr{\left( }

\def\cO{{\cal O}}
\def\dilog{{\rm{Li_2}}}
\def\im{{\rm{Im}}}
\newcommand\fverb{\setbox\pippobox=\hbox\bgroup\verb}
\newcommand\fverbdo{\egroup\medskip\noindent%
                        \fbox{\unhbox\pippobox}\ }
\newcommand\fverbit{\egroup\item[\fbox{\unhbox\pippobox}]}
\newbox\pippobox

\def\Li{{\rm Li}}

\def\nn{\nonumber}

\title{Scalar one-loop integrals for QCD}

\author{R. Keith Ellis, \\ 
Fermilab, Batavia, IL 60510, USA\\
Email:~\email{ellis@fnal.gov\\}}
 
\author{Giulia Zanderighi,\\
The Rudolf Peierls Centre for Theoretical Physics,\\
Department of Physics,University of Oxford,\\
1 Keble Road, OX1 3NP, Oxford, UK\\
Email:~\email{g.zanderighi1@physics.ox.ac.uk\\}}

\preprint{arXiv:0712.1851\\
FERMILAB-PUB-07-633-T\\
OUTP-07/16P}

\abstract{
We construct a basis set of infra-red and/or collinearly divergent
scalar one-loop integrals and give analytic formulas, for tadpole,
bubble, triangle and box integrals, regulating the divergences
(ultra-violet, infra-red or collinear) by regularization in
$D=4-2\epsilon$ dimensions. For scalar triangle integrals we give
results for our basis set containing 6 divergent integrals.  For
scalar box integrals we give results for our basis set containing 16
divergent integrals.  We provide analytic results for the 5 divergent
box integrals in the basis set which are missing in the literature.
Building on the work of van Oldenborgh, a general, publicly available
code has been constructed, which calculates both finite and divergent
one-loop integrals. The code returns the coefficients of
$1/\epsilon^2,1/\epsilon^1$ and $1/\epsilon^0$ as complex numbers for
an arbitrary tadpole, bubble, triangle or box integral.}


\begin{document} 

%\maketitle  IS IGNORED %%%%%%%%%%%

\section{Introduction} The advent of the LHC requires a concerted
effort to evaluate hard scattering processes at next-to-leading order
in QCD. This requires both the calculation of tree graphs, for the
leading order and real parton emission contributions, and the
calculation of one-loop diagrams, for the virtual contributions. For
most approaches to the calculation of one-loop amplitudes, the
knowledge of {\it scalar} one-loop integrals is sufficient.  For
integrals with all massive internal lines these integrals are all
known, both analytically~\cite{'t Hooft:1978xw,van
Oldenborgh:1989wn,Denner:1991qq} and numerically \cite{van
Oldenborgh:1990yc,Hahn:1998yk}.  This paper therefore concentrates on
integrals with some vanishing internal masses; these integrals can
contain infra-red and collinear singularities.

Many results have been presented in the literature for integrals with
infra-red divergences regulated by introducing a small mass $\lambda$
for the divergent lines, especially in the important paper of
Beenakker and Denner \cite{Beenakker:1988jr}. Regulation with a small
mass has been the method of choice in the calculation of electroweak
processes in which massless lines are relatively rare. In QCD
processes massless gluons are ubiquitous; in addition, light-quark
lines are often treated as massless in the high energy
limit. Consequently in QCD the method of choice for the regulation of
collinear and infra-red divergent integrals is dimensional
regularization.  Therefore the results given in this paper are
regulated dimensionally, although we shall exploit the relationship
between the two methods of regulation where appropriate.

Although some of the integral results presented in this paper are (to
the best of our knowledge) new, the majority are not.  However the
analysis of the basis set required for a complete treatment of
divergent box integrals at one loop is new. The results for the known
integrals are dispersed through the literature and we believe it will
be of use to collect the results in one place.  Note that the general
box integral in $D$-dimensions has been calculated in
ref.~\cite{Fleischer:2003rm}. However the results of that reference
require considerable further manipulation before they can be used in
practical calculations.

Given the complete set of tadpole, bubble, triangle and box integrals
one can construct the scalar integrals for diagrams with greater than
four legs~\cite{Melrose,van Neerven:1983vr,Bern:1992em}.  Thus a
scalar pentagon in $D$ dimensions, $I_5^D$, can be written as a sum of
the five box diagrams obtainable by removing one propagator if we
neglect terms of order $\epsilon$
\beq
I_5^{D} = \sum_{i=1}^5  c_i I_4^{D\; (i)} + \cO(\epsilon)\,.
\eeq
The general one loop $N$-point integral in $D=4-2\e$ dimensions for $N
\ge 6$ can be recursively obtained as a linear combination of pentagon
integrals~\cite{Melrose,van Neerven:1983vr} 
provided that the external momenta are
restricted to four dimensions\footnote{Relations which have the same structure
as Eq.~(\ref{NtoNminus1}) can also be derived, without the restriction that 
the external momenta lie in four dimensions. For the details of these relations 
we refer the reader to refs. \cite{Bern:1993kr,Binoth:1999sp,Duplancic:2003tv,Giele:2004iy}.}.
\beq \label{NtoNminus1}
I_N^{D} = \sum_{i=1}^N  d_i I_{N-1}^{D\; (i)}\,. 
\eeq
Thus for the purposes of next-to-leading order calculations, higher
point functions $N>4$ can be always reduced to sums of boxes.
Note that for the case $N \geq 7$ the coefficients 
$d_i$ in Eq.~(\ref{NtoNminus1}) are not unique.

\section{Definitions and notation} \subsection{Definition of
integrals}

\begin{figure}[h]
\begin{center}
\includegraphics[angle=270,scale=0.65]{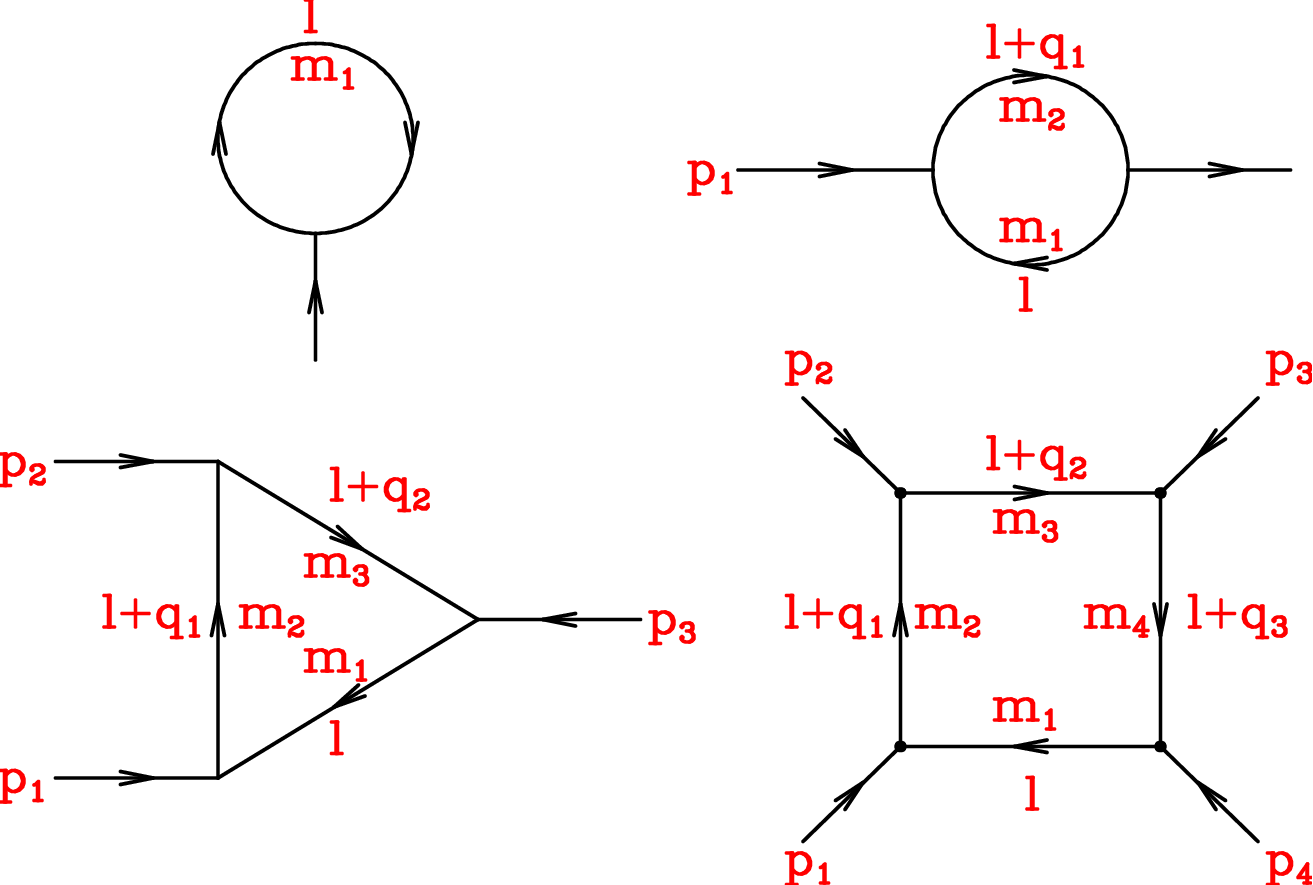}
\end{center}
\caption{The notation for the one-loop tadpole, bubble, triangle and box
integrals.}
\label{btbtfig}
\end{figure}

We work in the Bjorken-Drell metric so that
$l^2=l_0^2-l_1^2-l_2^2-l_3^2$. \ As illustrated in Fig.~\ref{btbtfig}
the definition of the integrals is as follows
\begin{eqnarray}
&& I^{D}_1(m_1^2)  =
 \frac{\mu^{4-D}}{i \pi^{\frac{D}{2}}\cG}\int d^D l \;
 \frac{1}{(l^2-m_1^2+i\varepsilon)}\,, \nn \\
&& I^{D}_2(p_1^2;m_1^2,m_2^2)  =
 \frac{\mu^{4-D}}{i \pi^{\frac{D}{2}}\cG}\int d^D l \;
 \frac{1}
{(l^2-m_1^2+i\varepsilon)
((l+q_1)^2-m_2^2+i\varepsilon)}\,,\nn \\
&& I^{D}_3(p_1^2,p_2^2,p_3^2;m_1^2,m_2^2,m_3^2)  =
\frac{\mu^{4-D}}{i \pi^{\frac{D}{2}}\cG}
\nn \\
&& \times \int d^D l \;
 \frac{1}
{(l^2-m_1^2+i\varepsilon)
((l+q_1)^2-m_2^2+i\varepsilon)
((l+q_2)^2-m_3^2+i\varepsilon)}\,,\nn \\
&&\nn \\
&&
I^{D}_4(p_1^2,p_2^2,p_3^2,p_4^2;s_{12},s_{23};m_1^2,m_2^2,m_3^2,m_4^2)
= 
\frac{\mu^{4-D}}{i \pi^{\frac{D}{2}}\cG}
\nn \\
&&
\times \int d^D l \;
 \frac{1}
{(l^2-m_1^2+i\varepsilon)
((l+q_1)^2-m_2^2+i\varepsilon)
((l+q_2)^2-m_3^2+i\varepsilon)
((l+q_3)^2-m_4^2+i\varepsilon)}\,, \nn \\ 
\end{eqnarray}
where $q_n\equiv \sum_{i=1}^n p_i$ and $q_0 = 0$ and $s_{ij}=
(p_i+p_j)^2$. For the purposes of this paper we take the masses in the
propagators to be real.  Near four dimensions we use $D=4-2 \e$.  (For
clarity the small imaginary part which fixes the analytic
continuations is specified by $+i\,\varepsilon$).  $\mu$ is a scale introduced so that the integrals
preserve their natural dimensions, despite excursions away from $D=4$.
We have removed the overall constant which occurs in $D$-dimensional integrals 
\beq
\cG\equiv\frac{\Gamma^2(1-\e)\Gamma(1+\e)}{\Gamma(1-2\e)} = 
\frac{1}{\Gamma(1-\e)} +{\cal O}(\e^3) =
1-\e \gamma+\e^2\Big[\frac{\gamma^2}{2}-\frac{\pi^2}{12}\Big]
+{\cal O}(\e^3)\,.
\eeq

The final results are given in terms of logarithms and dilogarithms.
The logarithm is defined to have a cut along the negative real axis.
The rule for the logarithm of a product is
\beqn
\ln(ab) &=& \ln a + \ln b +\eta(a,b)\,, \nn \\
\eta(a,b)&=& 2 \pi i \left[ 
\theta(-\im(a))\theta(-\im(b))\theta(\im(ab))
-\theta(\im(a))\theta(\im(b))\theta(-\im(ab))\right]\,.
\eeqn
So that
\beqn
\ln(ab) &=& \ln a + \ln b, \;\; \mbox{if}~\im(a)~\mbox{and}~\im(b)~\mbox{have different signs} \nn\,, \\
\ln\left(\frac{a}{b}\right) &=& \ln a - \ln b,\;\;\mbox{if}~\im(a)~\mbox{and}~\im(b) ~ \mbox{have the same sign}\,. 
\eeqn

The dilogarithm is defined as
\beqn
\dilog (x)&\equiv& -\int_0^x \frac{dz}{z} \ln(1-z)\\
&&=
\frac{x}{1^2}+\frac{x^2}{2^2}+\frac{x^3}{3^2}+\ldots ~\mbox{for}~ 
|{x}| \leq 1\,.
\eeqn
A number of the most commonly useful dilogarithm identities are given
in ref.~\cite{Lewin}. In addition to the common identities, we also
used an identity from ref.~\cite{Brandhuber:2004yw} in the calculation
of the new box integrals
\beqa
&&\dilog(1-aP^2)+\dilog(1-aQ^2)-\dilog(1-as)-\dilog(1-at)= \nn \\
&&
\dilog\Big(1-\frac{P^2}{s}\Big)
+\dilog\Big(1-\frac{P^2}{t}\Big)
+\dilog\Big(1-\frac{Q^2}{s}\Big)
+\dilog\Big(1-\frac{Q^2}{t}\Big) \nn \\
&-&\dilog\Big(1-\frac{P^2Q^2}{st}\Big)+\frac{1}{2} \ln^2\Big(\frac{s}{t}\Big)\; ,
\eeqa
which holds for $a=(P^2+Q^2-s-t)/(P^2 Q^2-st)$.

\subsection{Analytic continuation}
\label{sec:analytic}
The integrals given in sec.~\ref{sec:results} are calculated in the
spacelike region, $s_{ij}<0,p_i^2<0$. In this region, the denominator 
of the Feynman parameter integrals is positive definite and 
the $i\,\varepsilon$ prescription can be dropped. 
The analytic continuation is performed by restoring the
$i\,\varepsilon$
\beqa
\label{eq:cont}
p_{i}^2&\to& p_{i}^2+i\,\varepsilon\,,\nn \\ 
s_{ij}&\to & s_{ij}+i\,\varepsilon\,,\nn \\ 
m_{i}&\to & m_{i}-i\,\varepsilon\,. 
\eeqa
The one-loop integrals which we present are all expressed in terms of
two types of transcendental functions,
\beq
\label{eq:lnandli2}
\ln\left(\prod_{i=1}^{n} x_i\right)\,,\hspace{0.5cm}
{\rm and}\hspace{0.5cm}
\li\left(1-\prod_{i=1}^{n} x_i\right)\,.  
\eeq
The individual $x_i$ can be chosen such they that vary only on the
first Riemann sheet of the logarithm, $-\pi < {\rm arg}(x_i) < \pi$. A
complete specification of the continuation of terms of the form
eq.~(\ref{eq:lnandli2}) has been given in
ref.~\cite{Beenakker:1988jr}.
The continuation prescription for the logarithm is 
\beq \label{logcont}
\ln\left(\prod_{i=1}^{n} x_i\right) \to \sum_{i=1}^{n}
\ln\left(x_i\right)\,. 
\eeq
The continuation procedure for the dilogarithm is similar. For $\Big|
\prod_{i=1}^{n} x_i\Big| < 1$ we use~\cite{Beenakker:1988jr,Binoth:1999sp}
\beqa
\label{dilogcont}
\li\left(1-\prod_{i=1}^{n} x_i\right) &\to& 
\li\left(1-\prod_{i=1}^{n} x_i\right) 
+\ln\left(1-\prod_{i=1}^{n} x_i\right)
\Bigg[\ln\left(\prod_{i=1}^{n} x_i\right) -\sum_{i=1}^{n} \ln(x_i) \Bigg]
\nn \\
&=& \frac{\pi^2}{6} -\li\left(\prod_{i=1}^{n} x_i\right)
-\ln\left(1-\prod_{i=1}^{n} x_i\right)
\sum_{i=1}^{n} \ln(x_i) 
\,. 
\eeqa
If $\Big| \prod_{i=1}^{n} x_i\Big| > 1 $ it is expedient to make the
transformation
\beq
\li\left(1-\prod_{i=1}^{n} x_i\right) =- \li\left(1-\frac{1}{
\prod_{i=1}^{n} x_i}\right) -\frac{1}{2} \ln^2\left(\prod_{i=1}^{n} x_i\right)\,. 
\eeq
We can then continue the resulting expression using
eqs.~(\ref{logcont},~\ref{dilogcont}) as before.  The continuation
procedure given in ref.~\cite{van Hameren:2005ed} can be shown to be
equivalent to the above.

\section{Basis set of soft and collinear divergent integrals} After
Feynman parametrization and integration over $d^Dl$, we have for the
triangle and box integrals
\begin{eqnarray}
\label{eq:I3}
&& I^{D}_3(p_1^2,p_2^2,p_3^2;m_1^2,m_2^2,m_3^2) 
 = -\frac{\mu^{2 \e}\Gamma(1+\e)}{\cG}
\prod_{i=1}^3
\int_0^1 \!\!\! 
d a_k  \; \frac{\delta(1-\sum_k a_k)}
{\Big[\sum_{i,j} a_i a_j Y_{ij} -i \varepsilon \Big]^{1 + \e}}\,, \\ 
\label{eq:I4}
&& I^{D}_4(p_1^2,p_2^2,p_3^2,p_4^2;s_{12},s_{23};m_1^2,m_2^2,m_3^2,m_4^2) 
 = \frac{\mu^{2 \e}\Gamma(2+\e)}{\cG}
\prod_{i=1}^4
\int_0^1 \!\!\! 
%\int_{0\leq a_i \leq 1}\; 
d a_k  \; \frac{\delta(1-\sum_k a_k)}
{\Big[\sum_{i,j} a_i a_j Y_{ij} -i \varepsilon \Big]^{2 + \e}}\nonumber\,, \\
\end{eqnarray}
where $Y$ is the so-called modified Cayley matrix
\beq \label{Cayley}
 Y_{ij}\equiv \frac{1}{2} \Big[ m_i^2+m_j^2-(q_{i-1}-q_{j-1})^2\Big]\,.
\eeq
\subsection{Landau conditions}
\label{sec:landau}
The necessary conditions for eqs.~(\ref{eq:I3},\ref{eq:I4}) to contain
a singularity are due to Landau~\cite{Landau:1959,Eden:1966}.  If we
introduce the bilinear form $D$ derived from the modified Cayley
matrix,
\beq
D = \sum_{i,j} a_i a_j Y_{ij}\, ,
\eeq
eqs.~(\ref{eq:I3}, \ref{eq:I4}) contain singularities if $D=0$ and one
of the following conditions is satisfied for all values of $j$
\beqn \label{eq:Landau}
\mbox{\bf either}~a_j=0~\mbox{\bf or}~ \frac{\partial D} {\partial a_j}=0\, .
\eeqn

In general we have two classes of solutions of eq.~(\ref{eq:Landau}).
In the first class we have solutions for the $a_i$ which are implicit
functions of the masses and external momenta. Both physical and
anomalous thresholds fall into this class. The leading Landau
singularity occurs when $\frac{\partial D} {\partial a_j}=0$ is
satisfied for all $j$.  Such solutions of the Landau equations will
lead to divergences at the kinematic points where there are anomalous
and physical thresholds.

The second class of solution, which is of interest here, is 
the case where the external virtualities and internal masses have
fixed values and the Landau conditions have solutions for arbitrary
values of the other external invariants, $s_{ij}$.  Only these
solutions will lead to soft and collinear divergences which are
relevant for next-to-leading order calculations.

\begin{figure}[h]
\begin{center}
\includegraphics[angle=270,scale=0.65]{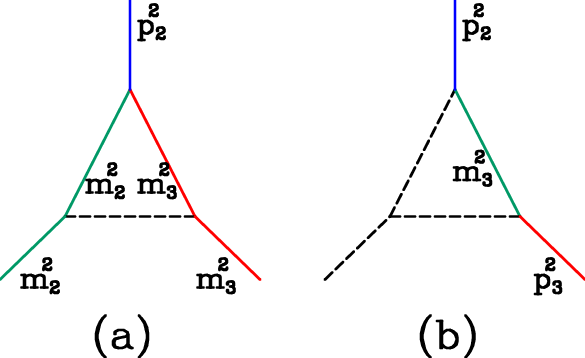}
\end{center}
\caption{Examples of triangle diagrams with divergences.}
\label{landau}
\end{figure}

As an example we consider the triangle shown in Fig.~(\ref{landau}a)
which contains a soft singularity. In this case the denominator is
given by
\beq
D=(m_2^2+m_3^2-\pd^2) a_2 a_3 + m_2^2 a_2^2+ m_3^2 a_3^2\,.
\eeq
This expression satisfies the Landau conditions for $a_2=a_3=0$ and
$a_1$ arbitrary.  A second example is the triangle shown in
Fig.~(\ref{landau}b) which contains a collinear singularity.  In this
case the denominator reads
\beq
D=(m_3^2-\pd^2) a_2 a_3 + (m_3^2-\pt^2) a_1 a_3 + m_3^2 a_3^2\,,
\eeq
which satisfies the Landau conditions for $a_3=0$ and $a_1,a_2$
arbitrary.

\subsection{Soft and collinear divergences}
From the Landau conditions it follows that a necessary condition for a
soft or collinear singularity is that for at least one value of the
index $i$~\cite{Kinoshita:1962ur}
\beq \label{soft}
Y_{i+1\; i+1}=Y_{i+1\; i+2}=Y_{i+1 \; i}=0  \;, \hspace{0.5cm} {\rm
soft\,\, singularity}\,, 
\eeq
\beq \label{collinear}
Y_{i\; i}=Y_{i+1\; i+1}=Y_{i \; i+1}=0 \;, \hspace{.5cm} {\rm
collinear\,\, singularity}\,. 
\eeq
The indices in eqs.~(\ref{soft}, \ref{collinear}) should be
interpreted $\mbox{mod}~N$, where $N$ is the number of external legs.
Thus the structure of the Cayley matrices for integrals having a soft
or collinear divergence is as follows
\beq
Y_{\rm soft}=
\left(\begin{array}{cccc}
\ldots & 0 & \ldots & \ldots \\
0 & 0 & 0& \ldots \\
\ldots & 0 & \ldots &\ldots\\
\ldots & \ldots & \ldots &\ldots \\
\end{array}\right),
\;\;\;\;
Y_{\rm collinear}=
\left(\begin{array}{cccc}
\ldots & \ldots & \ldots & \ldots \\
\ldots & 0 & 0& \ldots \\
\ldots & 0 & 0 &\ldots\\
\ldots & \ldots & \ldots &\ldots \\
\end{array}\right)\,.
\eeq
In order to have a divergence, we must have at least one internal mass
equal to zero, i.e. at least one vanishing diagonal element of $Y$.

\subsection{Basis set for triangle integrals}
\begin{figure}
\begin{center}
\includegraphics[angle=270,scale=0.7]{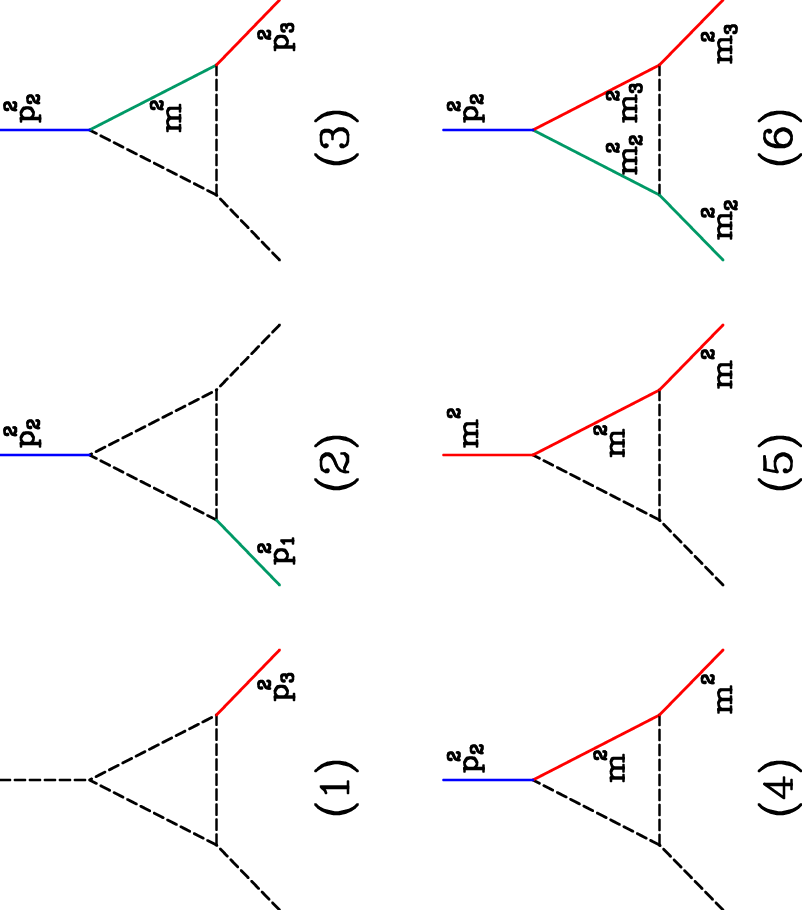}
\end{center}
\caption{The six divergent triangle integrals.
Lines with a zero internal mass $m_i=0$ (for internal lines) or a zero
virtuality, $p_i^2=0$, (for external lines) are shown dashed and are
unlabelled.  Solid lines have a non-zero internal mass, or a non-zero
virtuality. Lines with the same color have the same internal mass
and/or virtuality.}
\label{trifig}
\end{figure}
We list here the basis set of six divergent triangle integrals as
shown in Fig.~\ref{trifig}. All other divergent triangle integrals
can be derived from this set. First we have two integrals with no
internal masses,
\newcounter{soveenum}
\begin{enumerate}
\item  
{$I^{D}_3(0,0,p_3^2;0,0,0)$}
\item  
{$I^{D}_3(0,\pd^2,\pt^2,0;0,0,0)$}.
\setcounter{soveenum}{\value{enumi}}
\end{enumerate}
Second, we have three integrals with one massive internal line,
\begin{enumerate}
\setcounter{enumi}{\value{soveenum}}
\item
{$I_3^D(0,\pd^2,\pt^2;0,0,m^2)$}
\item
{$I_3^D(0,\pd^2,m^2;0,0,m^2)$}
\item
{$ I_3^D(0,m^2,m^2;0,0,m^2)$}.
\setcounter{soveenum}{\value{enumi}}
\end{enumerate}
Last, we have one integral with two massive internal lines,
\begin{enumerate}
\setcounter{enumi}{\value{soveenum}}
\item
{$I_3^D(m_2^2,\pd^2,m_3^2;0,m_2^2,m_3^2)$}.
\end{enumerate}
The set is in fact overcomplete since the second integral can be
obtained from the $m^2 \to 0$ limit of the third integral. However
from a numerical point of view it is expedient to categorize the
integrals by the number of vanishing internal masses and to treat the
two cases separately.

\subsection{Basis set for box integrals}

\begin{figure}
\begin{center}
\includegraphics[angle=270,scale=0.81]{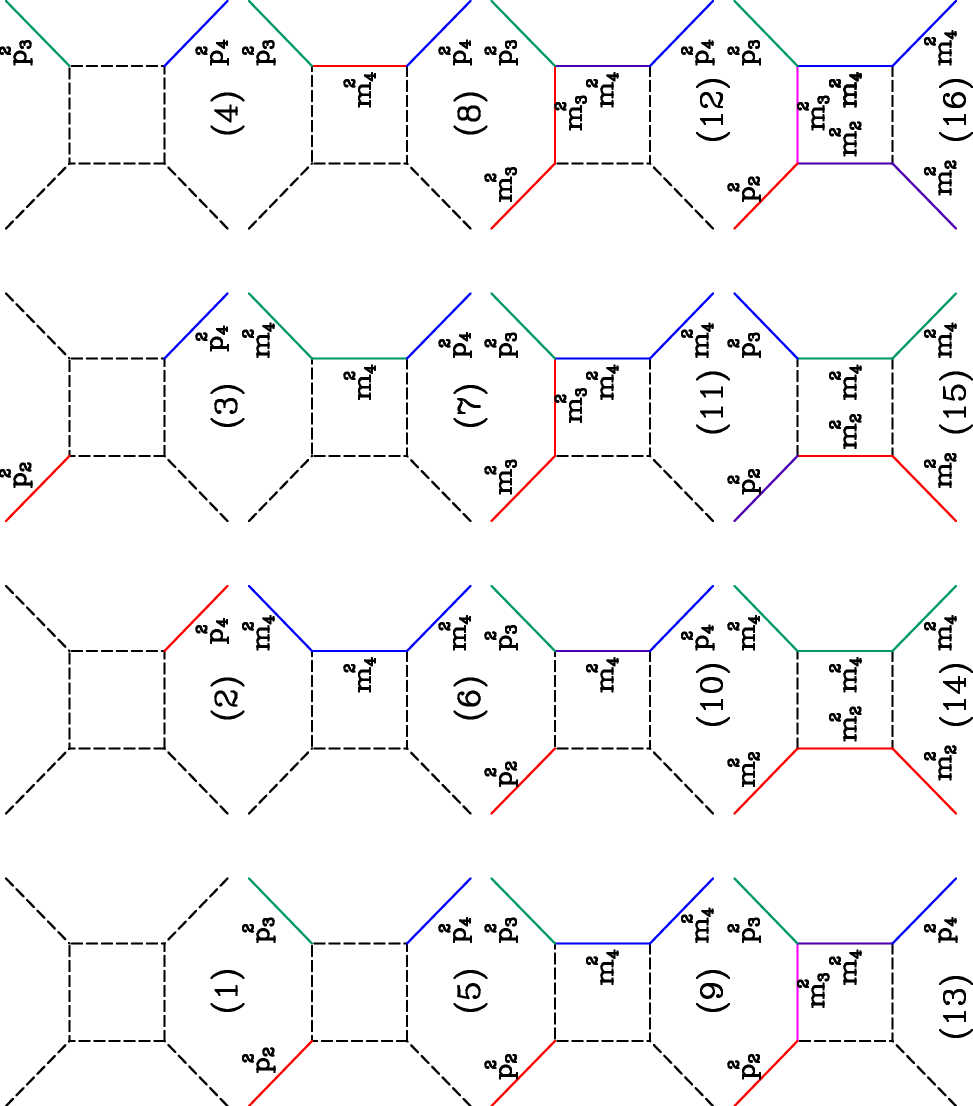}
\end{center}
\caption{The sixteen divergent box integrals.
Lines with a zero internal mass $m_i=0$ (for internal lines) or a zero
virtuality, $p_i^2=0$, (for external lines) are shown dashed and are
unlabelled. Solid lines have a non-zero internal mass, or a non-zero
virtuality. Lines with the same color have the same internal mass or
virtuality.}
\label{boxfig}
\end{figure}

\begin{figure}
\begin{center}
\includegraphics[angle=270,scale=0.7]{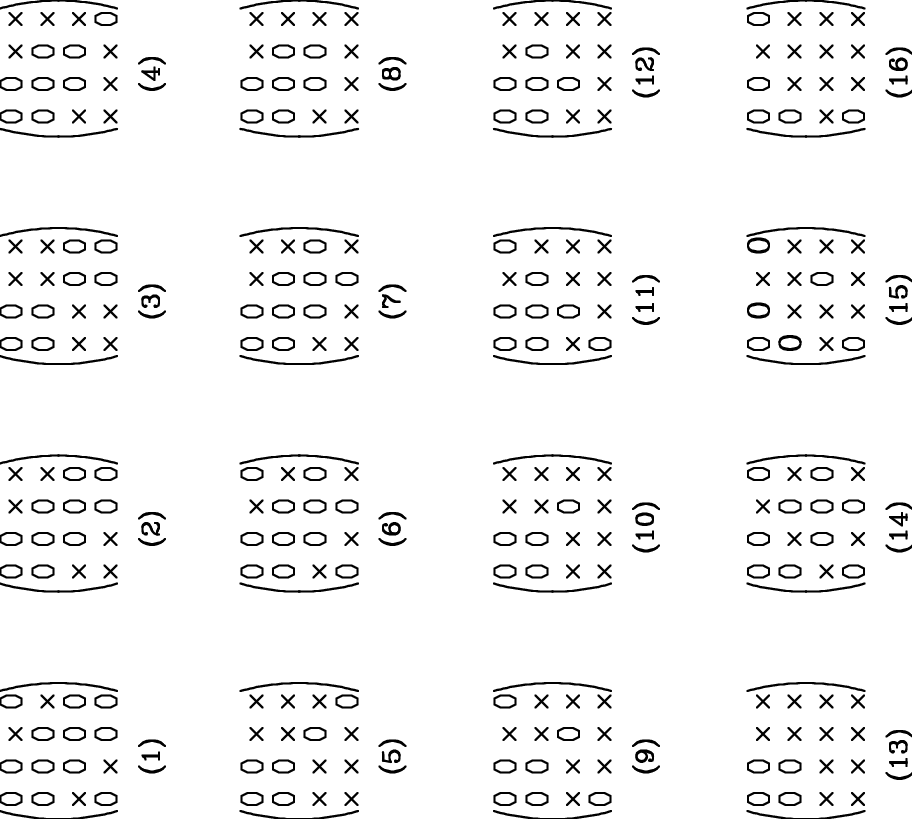}
\end{center}
\caption{The structure of the modified Cayley determinant 
for the sixteen divergent box integrals.}
\label{Cayleyfig}
\end{figure}

In this section we demonstrate that a basis set sufficient to describe
all box integrals with collinear or soft singularities can be
constructed from sixteen integrals, illustrated in Fig.~\ref{boxfig}.
All other divergent box integrals can be derived from this
set.~\footnote{As discussed in sec.~\ref{sec:landau}, at specific
kinematic points there are threshold singularities derivable from the
Landau conditions which can lead to singular integrals not derivable
from our basis set.}  The integrals are characterized by the number of
internal masses which are equal to zero. Each one of these divergent
integrals has a characteristic modified Cayley determinant as shown in
Fig.~\ref{Cayleyfig} satisfying the conditions of eqs.~(\ref{soft},
\ref{collinear}).

\subsubsection{Integrals with no internal masses}
There are five integrals with no internal masses,  
\newcounter{saveenum}
\begin{enumerate}
\item  {$I^D_4(0,0,0,0;s_{12},s_{23};0,0,0,0)$}
\item  {$I^D_4(0,0,0,p_4^2;s_{12},s_{23};0,0,0,0)$}
\item  {$I^D_4(0,p_2^2,0,p_4^2;s_{12},s_{23};0,0,0,0)$} %, "two opposite offshellness (easy)"
\item  {$I^D_4(0,0,p_3^2,p_4^2;s_{12},s_{23};0,0,0,0)$} %, "two adjacent offshellness (hard)"
\item  {$I^D_4(0,p_2^2,p_3^2,p_4^2;s_{12},s_{23};0,0,0,0)$}. 
\setcounter{saveenum}{\value{enumi}}
\end{enumerate}
This set of integrals is sufficient to describe all the integrals with
massless internal lines, but it is overcomplete since in
ref.~\cite{Duplancic:2000sk} an expression is given for box integral
3, which can also give results for integrals 1 and 2 by taking the
appropriate limit. However, as before, for reasons of numerical
expediency it is convenient to retain the overcomplete basis.

\subsubsection{Integrals with one non-zero internal mass}
If we have one non-zero internal mass we can take this without loss of
generality to be the last one ($m_4$).  In this case the modified
Cayley matrix, eq.~(\ref{Cayley}) is
\begin{eqnarray}
Y &=& \left(\begin{array}{cccc}
0                        &  -\frac{1}{2}p_1^2& -\frac{1}{2}s_{12} & \frac{1}{2}m_4^2-\frac{1}{2}p_4^2\\
-\frac{1}{2}p_1^2        &   0                            & -\frac{1}{2}p_2^2  &\frac{1}{2}m_4^2-\frac{1}{2}s_{23}\\
-\frac{1}{2}s_{12}       & -\frac{1}{2}p_2^2              & 0 &\frac{1}{2} m_4 ^2 -\frac{1}{2} p_3^2 \\
\frac{1}{2}m_4^2-\frac{1}{2}p_4^2 & \frac{1}{2}m_4^2-\frac{1}{2}s_{23}             & \frac{1}{2} m_4 ^2-\frac{1}{2} p_3^2 & m_4^2\\
\end{array} \right)\,.
\end{eqnarray}
With $s_{12},s_{23}$ fixed we can apply four conditions to potentially create
a soft or collinear divergence, namely 
\beq \label{fourconditions}
p_1^2=0,\; p_2^2=0,\; p_3^2=m_4^2,\; p_4^2=m_4^2.
\eeq
However performing the interchange $\pu^2 \leftrightarrow \pd^2,\pt^2
\leftrightarrow \pq^2$, (with $m_4$ fixed) corresponds to a
relabelling of the diagram.  In addition setting either $p_3^2=m_4^2$
without setting $\pd^2=0$, or $p_4^2=m_4^2$ without setting $\pu^2=0$
does not lead to a divergence.  If we denote the application of the
four conditions, eq.~(\ref{fourconditions}) on
$\pu^2,\pd^2,\pt^2,\pq^2$ by $(i,j,k,l)$ we have following 15 cases:
\begin{eqnarray} \label{fivelines}
&6.)\;\;\;\;\;\;& (1,2,3,4) \nonumber \\
&7.)\;\;\;\;\;\;& (1,2,3)\equiv (1,2,4) \nonumber \\
&8.)\;\;\;\;\;\;& (1,2) \nonumber \\
&9.)\;\;\;\;\;\;& (1,4)\equiv (2,3)\equiv (1,3,4) \equiv (2,3,4) \nonumber \\
&10.)\;\;\;\;\;\;&(1) \equiv (2) \equiv (1,3) \equiv (2,4) \nonumber \\
&\;\;\;\;\;\;&(3) \equiv (4) \equiv (3,4) \equiv \mbox{finite}\,.
\end{eqnarray}
To be quite explicit, the notation $(1,2,3)$ corresponds to setting
$p_1^2=0,p_2^2=0$ and $p_3^2=m_4^2$, etc.  The first five lines in
eq.~(\ref{fivelines}) correspond to the integrals 6-10 given below,
\begin{enumerate}
\setcounter{enumi}{\value{saveenum}}
\item {$I^{\{D=4-2 \epsilon\}}_4(0,0,m^2,m^2;s_{12},s_{23};0,0,0,m^2)$}
\item {$I^{\{D=4-2 \epsilon\}}_4(0,0,m^2,\pq^2;s_{12},s_{23};0,0,0,m^2)$}
\item {$I^{\{D=4-2 \epsilon\}}_4(0,0,\pt^2,\pq^2;s_{12},s_{23};0,0,0,m^2)$}
\item {$I^{\{D=4-2 \epsilon\}}_4(0,\pd^2,\pt^2,m^2;s_{12},s_{23};0,0,0,m^2)$}
\item {$I^{\{D=4-2 \epsilon\}}_4(0,\pd^2,\pt^2,\pq^2;s_{12},s_{23};0,0,0,m^2)$}.
\setcounter{saveenum}{\value{enumi}}
\end{enumerate}
The last case in eq.~(\ref{fivelines}) does not lead to a divergent integral.

\subsubsection{Integrals with two adjacent internal masses}
Without loss of generality we can take the two non-zero adjacent internal
masses to be $m_3$ and $m_4$.  In this case the modified Cayley matrix
is
\begin{eqnarray}
Y &=& \left(\begin{array}{cccc}
0                        &  -\frac{1}{2}p_1^2& \frac{1}{2}m_3^2 -\frac{1}{2}s_{12} & \frac{1}{2}m_4^2-\frac{1}{2}p_4^2\\
-\frac{1}{2}p_1^2        &   0                            & \frac{1}{2}m_3^2 -\frac{1}{2}p_2^2  &\frac{1}{2}m_4^2-\frac{1}{2}s_{23}\\
\frac{1}{2}m_3^2 -\frac{1}{2}s_{12}       & \frac{1}{2}m_3^2 -\frac{1}{2}p_2^2              & m_3^2 &\frac{1}{2}m_3^2 +\frac{1}{2} m_4 ^2 -\frac{1}{2} p_3^2 \\
\frac{1}{2}m_4^2-\frac{1}{2}p_4^2 & \frac{1}{2}m_4^2-\frac{1}{2}s_{23}             & \frac{1}{2}m_3^2 +\frac{1}{2} m_4 ^2-\frac{1}{2} p_3^2 & m_4^2\\
\end{array} \right)\,.
\end{eqnarray}
A necessary condition to have any divergence is $\pu^2=0$. This gives
the integral 13.  Applying either $\pd^2=m_3^2$ or $\pq^2=m_4^2$ gives
a pair of integrals related by relabelling, integral 12.  Applying
both $\pd^2=m_3^2$ and $\pq^2=m_4^2$ gives integral 11,
\begin{enumerate}
\setcounter{enumi}{\value{saveenum}}
\item {$I^{\{D=4-2 \epsilon\}}_4(0,m_3^2,\pt^2,m_4^2;s_{12},s_{23};0,0,m_3^2,m_4^2) $}
\item {$I^{\{D=4-2 \epsilon\}}_4(0,m_3^2,\pt^2,\pq^2;s_{12},s_{23};0,0,m_3^2,m_4^2) $}
\item {$I^{\{D=4-2 \epsilon\}}_4(0,\pd^2,\pt^2,\pq^2;s_{12},s_{23};0,0,m_3^2,m_4^2) $}.
\setcounter{saveenum}{\value{enumi}}
\end{enumerate}

\subsubsection{Integrals with two opposite internal masses}
Without loss of generality we can take the two non-zero opposite internal
masses to be $m_2$ and $m_4$.  In this case the modified Cayley matrix
is
\begin{eqnarray}
Y &=& \left(\begin{array}{cccc}
0                        &  \frac{1}{2}m_2^2-\frac{1}{2}p_1^2& -\frac{1}{2}s_{12} & \frac{1}{2}m_4^2-\frac{1}{2}p_4^2\\
\frac{1}{2}m_2^2-\frac{1}{2}p_1^2        &   m_2^2                           & \frac{1}{2}m_2^2-\frac{1}{2}p_2^2  &\frac{1}{2}m_2^2+\frac{1}{2}m_4^2-\frac{1}{2}s_{23}\\
-\frac{1}{2}s_{12}       & \frac{1}{2}m_2^2-\frac{1}{2}p_2^2              & 0 &\frac{1}{2} m_4 ^2 -\frac{1}{2} p_3^2 \\
\frac{1}{2}m_4^2-\frac{1}{2}p_4^2 & \frac{1}{2}m_2^2+\frac{1}{2}m_4^2-\frac{1}{2}s_{23}             & \frac{1}{2} m_4 ^2-\frac{1}{2} p_3^2 & m_4^2\\
\end{array} \right)\,.
\end{eqnarray}
Here we can only have a soft divergence since there is no pair of
adjacent zero internal masses.  Setting $\pu^2=m_2^2,\pq^2=m_4^2$ or
$\pd^2=m_2^2,\pt^2=m_4^2$ gives two integrals related by relabelling
(15). Setting both conditions gives integral 14, 
\begin{enumerate}
\setcounter{enumi}{\value{saveenum}}
\item {$I^{\{D=4-2 \epsilon\}}_4(m_2^2,m_2^2,m_4^2,m_4^2;s_{12},s_{23};0,m_2^2,0,m_4^2)$}
\item {$I^{\{D=4-2 \epsilon\}}_4(m_2^2,\pd^2,\pt^2,m_4^2;s_{12},s_{23};0,m_2^2,0,m_4^2)$}\,.
\setcounter{saveenum}{\value{enumi}}
\end{enumerate}

\subsubsection{Integral with three internal masses}
Without loss of generality we can take $m_1=0$.  With only one zero
mass, there can only be a soft singularity. This requires the two
adjacent external lines to satisfy the conditions,
$\pu^2=m_2^2,\pq^2=m_4^2$
\begin{enumerate}
\setcounter{enumi}{\value{saveenum}}
\item {$I_4^{D}(m_2^2,\pd^2,\pt^2,m_4^2;s_{12},s_{23};0,m_2^2,m_3^2,m_4^2)$}\,.
\end{enumerate}
The concludes the listing of our basis set of the divergent box integrals.

\section{Results for integrals}
\label{sec:results}
\subsection{Tadpole integrals}

The result for the tadpole integral is given by 
\beqn
I^{D}_1(m^2)  
&=& -\mu^{2\e} \; \Gamma(-1+\e) \;[m^2-i \varepsilon]^{1-\e} \nn \\
&=& \; m^2 \; \Big(\frac{\mu^2}{m^2-i\varepsilon}\Big)^\e
\left\{\frac{1}{\e}+1\right\} +\cO(\e) \,.
\eeqn
The $1/\e$ pole corresponds to an ultraviolet divergence and the
analytic continuation has been made explicit in this case.

\subsection{Bubble integrals}
Following 't Hooft and Veltman \cite{'t Hooft:1978xw}, 
the result for this bubble integral in our notation is,
\begin{eqnarray}
I^{D}_2(s;m_1^2,m_2^2)  
&=& \frac{\mu^{2\e} \Gamma(\e)}{\cG} \int_0^1 \; d\gamma \; 
[-\gamma (1-\gamma) s +\gamma m_2^2+(1-\gamma) m_1^2 -i \varepsilon]^{-\e}  \nonumber \\
&=& \mu^{2\e} %\Gamma(1+\e) 
\left\{ \frac{1}{\e}
 -\int_0^1 \; d\gamma \; \ln \big(-\gamma (1-\gamma) s +\gamma m_2^2+(1-\gamma) m_1^2 -i \varepsilon \big) \right\}
 +\cO(\e)\nonumber \\
&=& \mu^{2\e} %\Gamma(1+\e) 
\left\{ \frac{1}{\e} - \ln (s-i \varepsilon)
 -\int_0^1 \; d\gamma \; \ln \Big(\gamma^2-\gamma (1-\frac{m_2^2}{s}+\frac{m_1^2}{s})
+\frac{m_1^2}{s}- \frac{i \varepsilon}{s}\Big)
\right\} +\cO(\e)\nonumber \\
&=& \mu^{2\e} %\Gamma(1+\e) 
\left\{ \frac{1}{\e} +2 - \ln (s-i \varepsilon)
 +\sum_{i=1}^{2} \Big[ \gamma_i \ln \big(\frac{\gamma_i-1}{\gamma_i}\big) -\ln (\gamma_i-1) \Big]
\right\} +\cO(\e)\,,\nonumber \\
\end{eqnarray}
where $\gamma_{1,2}$ are the two roots of the quadratic equation,
\begin{equation}
\gamma_{1,2} = \frac{s-m_2^2+m_1^2\pm \sqrt{(s-m_2^2+m_1^2)^2
-4 s (m_1^2-i \varepsilon)}}{2 s}\,.
\end{equation}
The special limits for this integral are
\beqa
I_2^{D}(s;0,m^2)&=& \Big(\frac{\mu^2}{m^2} \Big)^{\e} 
\left\{ \frac{1}{\e}+2 +\frac{m^2-s}{s} \ln\Big( \frac{m^2-s -i
\varepsilon}{m^2}\Big)\right\}+\cO(\e)\,,\nn \\
%\eeq
%\beq
I_2^{D}(s;0,0)&=& \Big(\frac{\mu^2}{-s-i \varepsilon} \Big)^{\e} 
\left\{ \frac{1}{\e}+2 \right\}+{\cal O}(\e)\,.
\eeqa
As before, the $1/\e$ pole corresponds to an ultraviolet divergence
and the analytic continuation has been made explicit in these cases.

\subsection{Divergent triangle integrals} In this section we give the
explicit results for the six divergent triangles. The results for the
triangles have been presented already by many authors and are given
here only for completeness.  The results are reported in the spacelike
region below all thresholds. The analytic continuation is performed
using the prescription given in sec.~\ref{sec:analytic}.
Each expression for a triangle integral stands for the 6 different
labellings of the triangle obtained by repeated application of the
following identities
\beqn
I^D_3(\pu^2,\pd^2,\pt^2;m_1^2,m_2^2,m_3^2)
= I^D_3(\pd^2,\pt^2,\pu^2;m_2^2,m_3^2,m_1^2)\,, \nn \\
I_3^D (\pu^2, \pd^2, \pt^2; m_1^2, m_2^2, m_3^2)  =   
I_3^D (\pu^2, \pt^2, \pd^2; m_2^2, m_1^2, m_3^2) \,.\nn 
\eeqn

\subsubsection{Triangle 1: $I_3^{D}(0,0,p_3^2;0,0,0)$}
\beqn
I_3^{D}(0,0,p_3^2;0,0,0)&=&
\frac{\mu^{2 \e}}{\e^2} \left\{\frac{(-p_3^2)^{-\e}}{p_3^2}\right\}\nn \\
&=&\frac{1}{p_3^2} \left\{\frac{1}{\e^2}
-\frac{1}{\e} \ln \Big( \frac{-p_3^2}{\mu^2}
\Big)
+\frac{1}{2} \ln^2 \Big( \frac{-p_3^2}{\mu^2}\Big)\right\}+\cO(\e).
\eeqn

\subsubsection{Triangle 2: $I_3^{D}(0,\pd^2,\pt^2;0,0,0)$}
\beqa
\label{eq:tri2}
&&I_3^{D}(0,\pd^2,\pt^2,0;0,0,0)=
\frac{\mu^{2 \e}}{\e^2} \Bigg\{\frac{(-\pd^2)^{-\e}-(-\pt^2)^{-\e}}{\pd^2-\pt^2
}\Bigg\}\nn \\
&=&\frac{1}{{\pd^2-\pt^2}} \Bigg\{\frac{1}{\e} \ln\Big(\frac{-\pt^2}{-\pd^2}\Big)
+\frac{1}{2}\Big[
\ln^2\Big(\frac{-\pd^2}{\mu^2}\Big)
    -\ln^2\Big(\frac{-\pt^2}{\mu^2}\Big)\Big]\Bigg\}+\cO(\e)\,.
\eeqa
In the limit $p_2^2 \to p_3^2$ we define $r = (\pt^2-\pd^2)/\pd^2$ and
use following expansion 
\beq
I_3^{D}(0,\pd^2,\pt^2,0;0,0,0)=
 \frac{1}{\pd^2} \Bigg\{
-\frac{1}{\e}\left(1-\frac{r}{2}\right)
+\ln\left(\frac{-\pd^2}{\mu^2}\right)+\frac{r}{2}\left(1+\ln\left(\frac{-\pd^2}{\mu^2}\right)\right)
\Bigg\}+\cO(\e,r^2)\,.
\eeq

\subsubsection{Triangle 3: $I_3^{D}(0,\pd^2,\pt^2;0,0,m^2)$}
\beqa
\label{eq:tri3}
&&I_3^{D}(0,\pd^2,\pt^2;0,0,m^2)=\frac{1}{\pd^2-\pt^2 }
\left(\frac{\mu^2}{m^2}\right)^\e \Bigg\{\frac{1}{\e}
\ln\Big(\frac{m^2-\pt^2}{m^2-\pd^2}\Big) +\li\Big(\frac{\pd^2}{m^2}\Big)-\li\Big(\frac{\pt^2}{m^2}\Big) \nn \\
&+&\ln^2\Big(\frac{m^2-\pd^2}{m^2}\Big)
-\ln^2\Big(\frac{m^2-\pt^2}{m^2}\Big) \Bigg\}+O(\e)\,.
\eeqa

In the limit $p_2^2 \to p_3^2$ we define $r = (\pt^2-\pd^2)/(m^2-\pd^2)$ and
use following expansion 
\beqa
&&I_3^{D}(0,\pd^2,\pt^2,0;0,0,0)=\frac{1}{m^2-\pd^2}
 \Bigg\{
\left(1-\frac{r}{2}\right)
\left(\frac{1}{\e}-\ln\frac{m^2}{\mu^2}\right) 
-\frac{m^2+\pd^2}{\pd^2}\ln\left(\frac{m^2-\pd^2}{m^2}\right)\nn\\
&-& \frac{r}{2 \pd^2}\left[
\frac{m^4-2 \pd^2 m^2
-\pd^4}{\pd^2}\ln\left(\frac{m^2-\pd^2}{m^2}\right)
+m^2 +\pd^2
\right]
\Bigg\}+\cO(\e,r^2)\,.
\eeqa

Rewriting eq.~(\ref{eq:tri3}) in the following form makes the $m \to 0
$ limit and the agreement with eq.~(\ref{eq:tri2}) manifest
\beqa
&&I_3^{D}(0,\pd^2,\pt^2;0,0,m^2)=\frac{1}{\pd^2-\pt^2 } 
\Bigg\{ \frac{1}{\e} \ln\Big(\frac{m^2-\pt^2}{m^2-\pd^2}\Big) 
+\frac{1}{2}\Big[\ln^2\Big(\frac{-\pd^2}{\mu^2}\Big) 
-\ln^2\Big(\frac{-\pt^2}{\mu^2}\Big)\Big] \nn\\
&+&\ln\Big(\frac{(m^2-\pd^2)}{-\pd^2}\Big)
 \ln\Big(\frac{-\pd^2(m^2-\pd^2)}{\mu^2 m^2}\Big)
- \ln\Big(\frac{(m^2-\pt^2)}{-\pt^2}\Big)
 \ln\Big(\frac{-\pt^2(m^2-\pt^2)}{\mu^2 m^2}\Big)\nn\\
&-&\li\Big(\frac{m^2}{\pd^2}\Big)+\li\Big(\frac{m^2}{\pt^2}\Big) 
\Bigg\}+O(\e)\,.
\eeqa

\subsubsection{Triangle 4: $I_3^{D}(0,\pd^2,m^2;0,0,m^2)$}
\beqa
&&I_3^{D}(0,\pd^2,m^2;0,0,m^2)=\left(\frac{\mu^2}{m^2}\right)^\e 
\frac{1}{\pd^2-m^2} \nn \\
&\times & \left\{\frac{1}{2 \e^2}
 +\frac{1}{\e} \ln\left(\frac{m^2}{m^2-\pd^2}\right)
+\frac{\pi^2}{12}+\frac{1}{2}\ln^2\left(\frac{m^2}{m^2-\pd^2}\right)
-{\li}\left(\frac{-\pd^2}{m^2-\pd^2}\right)
\right\}+\cO(\e)\,.\nn \\ 
\eeqa

\subsubsection{Triangle 5: $I_3^{D}(0,m^2,m^2;0,0,m^2)$}
\begin{equation}
I_3^{D}(0,m^2,m^2;0,0,m^2)=\left(\frac{\mu^2}{m^2}\right)^\e 
\frac{1}{m^2} \Bigg(-\frac{1}{2 \e}+1 \Bigg)+\cO(\e)\,.
\end{equation}

\subsubsection{Triangle 6: $I^{D}_3(m_2^2,s,m_3^2;0,m_2^2,m_3^2)$}
The result for this triangle integral can be obtained from
ref.~\cite{Beenakker:1988jr}, eq.~(C3), by the normal replacement
rule \cite{Dittmaier:2003bc} which is true in the case of a soft
singularity 
\beq
\label{eq:rep}
\ln\lambda^2 \to
\frac{\cG}{\epsilon}+\ln\mu^2+\cO{(\e)}\,.
\eeq

For $s-(m_2-m_3)^2\neq 0$ we have
\beqn
&&I^{\{D=4\}}_3(m_2^2,s,m_3^2;0,m_2^2,m_3^2) =
\frac{x_s}{m_2 m_3 (1-x_s^2)} \nn \\ 
&&\times \Bigg\{\ln(x_s) \Big[-\frac{1}{\e} -\frac{1}{2} \ln(x_s)+2
\ln(1-x_s^2)
+\ln(\frac{m_1 m_3}{\mu^2})\Big] \nn \\
&-& \frac{\pi^2}{6} +\li(x_s^2)+\frac{1}{2} \ln^2\frac{m_2}{m_3}+\li(1-x_s\frac{m_2}{m_3})+\li(1-x_s\frac{m_3}{m_2})\Bigg\}
+\cO(\e)\,,
\eeqn
where $x_s=-K(s+i \varepsilon,m_2,m_3)$ and $K$ is given by
\beqn
\label{eq:Kdef}
K(z,m,m^\prime) &= \frac{1-\sqrt{1-4 m m^\prime /[z-(m-m^\prime)^2]}}{1+\sqrt{1-4 m m^\prime /[z-(m-m^\prime)^2]}}
\;\; &z \neq (m-m^\prime)^2 \nn \\
K(z,m,m^\prime) &= -1\;\; &z = (m-m^\prime)^2\,. 
\eeqn
For $s-(m_2-m_3)^2=0$ this becomes
\beqn
&&I^{\{D=4\}}_3(m_2^2,s,m_3^2;0,m_2^2,m_3^2) = \frac{1}{2 m_2 m_3} \nn
\\ 
&&\times \Bigg\{ \frac{1}{\e}+\ln\left(\frac{\mu^2}{m_1 m_3}\right)-2
-\frac{m_3+m_2}{m_3-m_2}\ln\left(\frac{m_2}{m_3}\right)\Bigg\}+\cO(\e)\,.
\eeqn

\subsection{Divergent box integrals}
\label{sec:boxes}

In this section we give the explicit results for the sixteen divergent
boxes. 
The results are reported in the spacelike region below all
thresholds. The analytic continuation is performed using the
prescription given in sec.~\ref{sec:analytic}. Each expression for a
box integral stands for the 8 different labellings of the box obtained
by repeated application of the following identities
\beqn
I^D_4(\pu^2,\pd^2,\pt^2,\pq^2,\sud,\sdt,m_1^2,m_2^2,m_3^2,m_4^2)
= I^D_4(\pd^2,\pt^2,\pq^2,\pu^2,\sdt,\sud,m_2^2,m_3^2,m_4^2,m_1^2)\,, \nn \\
I^D_4(\pu^2,\pd^2,\pt^2,\pq^2,\sud,\sdt,m_1^2,m_2^2,m_3^2,m_4^2)
=I^D_4(\pq^2,\pt^2,\pd^2,\pu^2,\sud,\sdt,m_1^2,m_4^2,m_3^2,m_2^2)\,.\nn\\ 
\eeqn
Where we have found the integrals in the literature we give
references. To the best of our knowledge the results for boxes
9, 10, 11, 12, 13 are new. 

\subsubsection{Box 1: $I_4^{D}(0,0,0,0;s_{12},s_{23};0,0,0,0)$}
\beqa
&&I_4^{D}(0,0,0,0;\sud,\sdt;0,0,0,0)
=\frac{\mu^{2 \e}}{\sud \sdt} \nonumber \\
&\times& \left\{\frac{2}{\e^2} 
\Big((-\sud )^{-\e}
     +(-\sdt)^{-\e}\Big)
-\ln^2\Big(\frac{-\sud}{-\sdt}\Big) 
- \pi^2 \right\}+{\cal O}(\e)\,.
\eeqa
This result is taken from \cite{Bern:1993kr}.

\subsubsection{Box 2: $I_4^{D}(0,0,0,\pq^2;s_{12},s_{23};0,0,0,0)$}
\beqa
&&I_4^{D}(0,0,0,\pq^2;s_{12},s_{23};0,0,0,0)
=\frac{\mu^{2 \e}}{\sud \sdt} \nonumber \\
&\times& \left\{\frac{2}{\e^2} \Big((-\sud)^{-\e}+(-\sdt)^{-\e}-(-\pq^2)^{-\e}\Big)
 -2 \,\li(1-\frac{\pq^2}{\sud}) - 2 \, \li(1-\frac{\pq^2}{\sdt})\right. \nonumber \\
  &-&\left. \ln^2\Big(\frac{-\sud}{-\sdt}\Big)-\frac{\pi^2}{3}\right\}+{\cal O}(\e)\,.
\eeqa
This integral is given in \cite{Ellis:1980wv,Bern:1993kr}. 
An alternative formulation with three dilogarithms is given in
\cite{Duplancic:2000sk}.

\subsubsection{Box 3: $I_4^{D}(0,\pd^2,0,\pq^2;s_{12},s_{23};0,0,0,0)$}
\beqa \label{box3}
&&I_4^{D}(0,\pd^2,0,\pq^2;s_{12},s_{23};0,0,0,0)
= \frac{\mu^{2 \e}}{\sdt\sud- \pd^2 \pq^2} \nonumber \\
& \times &
\left\{\frac{2}{\e^2} 
\left((-\sud)^{-\e}+(-\sdt)^{-\e}
      -(-\pd^2)^{-\e}-(-\pq^2)^{-\e}\right)\right. \nonumber \\
 &-&
     2\,\li\left(1-\frac{\pd^2}{ \sud}\right)                      
    -2\,\li\left(1-\frac{\pd^2}{ \sdt}\right)                      
    -2\,\li\left(1-\frac{\pq^2}{ \sud}\right)                      
    -2\,\li\left(1-\frac{\pq^2}{ \sdt}\right) \nonumber \\
    &+&
\left.
     2\, \li\left(1-\frac{\pd^2 \pq^2 }{ \sud\sdt}\right)
    -\ln^2\left(\frac{-\sud}{-\sdt}\right) \right\}+{\cal O}(\e)\,.
\eeqa
This result is taken from \cite{Bern:1993kr}.
As for all the integrals, the analytic continuation of this result
follows the procedure detailed in sec.~\ref{sec:analytic}.  An
alternative form in which the analytic continuation is manifest is
given in ref.~\cite{Duplancic:2000sk}.

When the denominator in the overall factor in eq.~(\ref{box3}) 
becomes small and the entries in the two pairs 
$(\sud $, $\sdt)$  and $(\pd^2,\pq^2)$ have the opposite sign
we can expand in $r=1-\frac{\pd^2 \pq^2 }{\sud \sdt}$
\beqa 
&&I_4^{D}(0,\pd^2,0,\pq^2;s_{12},s_{23};0,0,0,0)
=  \frac{1}{\sdt\sud} \nonumber \\
& \times &
\left\{-\frac{1}{\e} (2+r)
       +(2-\frac{1}{2} r)+(2+r) 
 \left(\ln\left(\frac{-\sud}{\mu^2}\right)
       +\ln\left(\frac{-\sdt}{-\pq^2}\right)\right)  \right. \nn \\
    &+& \left. 2\,\left[\Ll_0\left(\frac{\pq^2}{\sdt}\right)
    + \Ll_0\left(\frac{\pq^2}{\sud}\right)\right]                      
    +r \left[\Ll_1\left(\frac{\pq^2}{\sdt}\right)
        +\Ll_1\left(\frac{\pq^2}{\sud}\right)\right]\right\}+{\cal O}(\e,r^2)\,,
\eeqa
where
\beq \label{Ldefs}
\Ll_0(z)=\frac{\ln\left(z\right)}{1-z},\;\;\;\;
\Ll_1(z)=\frac{\Ll_0\left(z\right)+1}{1-z} 
\eeq
Thus in this region the residue of the overall pole 
at $\sdt\sud= \pd^2 \pq^2$
vanishes and we obtain a numerically stable expression.
The other region $\sud,\sdt>0$ and $\pd^2,\pq^2<0$ 
or vice versa ($\sud,\sdt<0$ and $\pd^2,\pq^2>0$) is the region of the Landau
pole which gives a large contribution to the imaginary part.

\subsubsection{Box 4: $I_4^{D}(0,0,\pt^2,\pq^2;s_{12},s_{23};0,0,0,0)$}
\beqa
\label{eq:box4}
&&I_4^{D}(0,0,\pt^2,\pq^2;s_{12},s_{23};0,0,0,0)=
\frac{\mu^{2 \e}}{\sud \sdt} \nonumber \\
  &\times& \left\{\frac{2}{\e^2} \Big((-\sud)^{-\e}+(-\sdt)^{-\e}
-(-\pt^2)^{-\e}-(-\pq^2)^{-\e})
  +\frac{1}{\e^2} \Big((-\pt^2)^{-\e}
  (-\pq^2)^{-\e}\Big)/(-\sud)^{-\e}\right. \nonumber \\ & &-\left. 2\,
  \li\left(1-\frac{\pt^2}{ \sdt}\right)
     -2\, \li\left(1-\frac{\pq^2}{ \sdt}\right)
     -\ln^2\left(\frac{-\sud}{-\sdt}\right) \right\}+{\cal O}(\e)\,.
\eeqa
This result is taken from ref. \cite{Bern:1993kr}. (See also
ref. \cite{Duplancic:2000sk}). 

\subsubsection{Box 5: $I_4^{D}(0,\pd^2,\pt^2,\pq^2;s_{12},s_{23};0,0,0,0)$}
\beqa \label{box5}
&&I_4^{D}(0,\pd^2,\pt^2,\pq^2;s_{12},s_{23};0,0,0,0)=\frac{\mu^{2 \e}}
{(\sdt\sud- \pd^2 \pq^2)}\nonumber \\
  &\times& \left\{\frac{2}{\e^2} \Big((-\sud)^{-\e}+(-\sdt)^{-\e}
-(-\pd^2)^{-\e}-(-\pt^2)^{-\e}-(-\pq^2)^{-\e}) \right. \nonumber \\
  &+&\frac{1}{\e^2} \Big((-\pd^2)^{-\e} (-\pt^2)^{-\e}\Big)/(-\sdt)^{-\e} 
    +\frac{1}{\e^2} \Big((-\pt^2)^{-\e} (-\pq^2)^{-\e}\Big)/(-\sud)^{-\e} 
\\
  &-& \left. 2\,\li\left(1-\frac{\pd^2}{ \sud}\right)
     -2\,\li\left(1-\frac{\pq^2}{ \sdt}\right)
     +2\,\li\left(1-\frac{\pd^2\pq^2}{\sud \sdt}\right)
-\ln^2 \Big(\frac{-\sud}{-\sdt}\Big) \right\}+{\cal O}(\e)\,.\nonumber
\eeqa
This result is taken from \cite{Bern:1993kr}.
As for all the integrals, the analytic continuation of this result
follows the procedure detailed in sec.~\ref{sec:analytic}.  An
alternative form in which the analytic continuation is manifest is
given in ref.~\cite{Duplancic:2000sk}.

When the denominator in the overall factor in eq.~(\ref{box5}) 
becomes small and the entries in the two pairs 
$(\sud $, $\sdt)$  and $(\pd^2,\pq^2)$ have the opposite sign
we can expand in $r=1-\frac{\pd^2 \pq^2 }{\sud \sdt}$
\beqa
&&I_4^{D}(0,\pd^2,\pt^2,\pq^2;s_{12},s_{23};0,0,0,0) = \frac{1}
{\sdt\sud}\nonumber \\
  &\times& \left\{-\frac{1}{\e} \left(1+\frac{1}{2}r\right)
-(1+\frac{1}{2}r) \left[\ln\left(\frac{\mu^2}{-\sud}\right)
+\ln\left(\frac{-\pt^2}{-\sdt}\right)-2
-(1+\frac{\pq^2}{\sdt}) \Ll_0\left(\frac{\pq^2}{\sdt}\right)\right] 
\right. \nn \\
 &+&\left. r\left[\Ll_1\left(\frac{\pq^2}{\sdt}\right)
                    -\Ll_0\left(\frac{\pq^2}{\sdt}\right)-1 \right]
\right\}+{\cal O}(\e,r^2)\, ,
\eeqa
with  $\Ll_0,\Ll_1$ as in eq.~(\ref{Ldefs}).

\subsubsection{Box 6: $I_4^{D}(0,0,m^2,m^2;s_{12},s_{23};0,0,0,m^2)$}
\beqa
&& I_4^{\{D=4-2 \e\}}(0,0,m^2,m^2;s_{12},s_{23};0,0,0,m^2)=
 -\frac{1}{s_{12} (m^2-s_{23})} \left( \frac{\mu^2}{m^2}\right)^\e \nonumber \\
&& \times \left\{
\frac{2}{\e^2}
 -\frac{1}{\e} \Big(2 \ln(\frac{m^2-s_{23}}{m^2})+\ln(\frac{-s_{12}}{m^2})\Big)
+2\ln(\frac{m^2-s_{23}}{m^2})\ln(\frac{-s_{12}}{m^2})-\frac{\pi^2}{2}\right\}
+\cO(\e)\,. 
\nn \\
\eeqa
The result for the real part of this integral in the region $s_{12}>0,
s_{23}<0$ is given in ref.~\cite{Beenakker:1988bq} eq.~(A4) (note
differing definition of $\e$).

\subsubsection{Box 7: $I_4^{D}(0,0,m^2,\pq^2;s_{12},s_{23};0,0,0,m^2)$}
\beqa
&& I_4^{D}(0,0,m^2,\pq^2;s_{12},s_{23};0,0,0,m^2)=
 \left( \frac{\mu^2}{m^2} \right)^{\e}
 \frac{1}{s_{12} (s_{23}-m^2)} \quad \nn\\
& \times  &\left\{ 
  \frac{3}{2}\frac{1}{\e^2}
  - \frac{1}{\e}
\bigg[ 2 \ln\bigg(1-\frac{ s_{23}}{m^2}\bigg) 
  + \ln\bigg(\frac{{}-{ s_{12}}\,}{m^2}\bigg) 
  - \ln\bigg(1-\frac{ \pq^2}{m^2}\bigg) \bigg]
\right.
\nn\\ && \qquad \left.
  - 2\, \Li\bigg( \frac{ s_{23}- \pq^2}{ s_{23}-m^2} \bigg)
  + 2 \ln\bigg(\frac{{}-{  s_{12}}\,}{m^2}\bigg) 
      \ln\bigg(1-\frac{ s_{23}}{m^2}\bigg) 
  - \ln^2\bigg(1-\frac{ \pq^2}{m^2}\bigg) 
  - \frac{5\pi^2}{12} \right\} + \cO(\e)\nn \\
\eeqa
This integral is obtained from eq.~(A4) (first equation) of
ref.~\cite{Beenakker:2002nc}.  (The real part was given earlier in
eq.~(6.75) of \cite{Hopker:1996sx}.)

\subsubsection{Box 8: $I_4^{D}(0,0,\pt^2,\pq^2; s_{12},s_{23};0,0,0,m^2)$}
\beqa
\label{eq:box8}
&&I_4^{D}(0,0,\pt^2,\pq^2;s_{12},s_{23};0,0,0,m^2) =
 \frac{1}{s_{12}(s_{23}-m^2)} \nn \\
&&\times 
 \left\{ \frac{1}{\epsilon^2}-\frac{1}{\epsilon}
 \Big[\ln\frac{-s_{12}}{\mu^2}+\ln\frac{(m^2-s_{23})^2}{(m^2-\pt^2)(m^2-\pq^2)}
  \Big] \right. \nonumber \\
 &&
 -2\; \li\lr 1-\frac{m^2-\pt^2}{m^2-s_{23}}\rr
 -2\; \li\lr 1-\frac{m^2-\pq^2}{m^2-s_{23}}\rr
 - \li\lr 1+\frac{(m^2-\pt^2)(m^2-\pq^2)}{s_{12} m^2}\rr \nonumber \\
 &&
 -\frac{\pi^2}{6}
 +\frac{1}{2}\ln^2\lr\frac{-s_{12}}{\mu^2}\rr
 -\frac{1}{2}\ln^2\lr\frac{-s_{12}}{m^2}\rr
 +2\ln\lr\frac{-s_{12}}{\mu^2}\rr\ln\lr\frac{m^2-s_{23}}{m^2}\rr \nonumber \\
 &&\left. 
  -\ln\lr\frac{m^2-\pt^2}{\mu^2}\rr\ln\lr\frac{m^2-\pt^2}{m^2}\rr
 -\ln\lr\frac{m^2-\pq^2}{\mu^2}\rr\ln\lr\frac{m^2-\pq^2}{m^2}\rr \right\} +  \cO(\e).
\eeqa

This integral was constructed from the expression eq.~(B6) of
ref.~\cite{Berger:2000iu}, which is valid for the real part in the
region $\sud > 0, \sdt <0$.

\subsubsection{Box 9: $I^{D}_4(0,p_2^2,p_3^2,m^2;s_{12},s_{23};0,0,0,m^2)$}
\beqa
&&I^{D}_4(0,p_2^2,p_3^2,m^2;s_{12},s_{23};0,0,0,m^2)
=\frac{1}{\sud (\sdt-m^2)}\left\{
       \frac{1}{2 \e^2}
       - \frac{1}{\e}
          \ln\Big(\frac{\sud}{\pd^2} \frac{(m^2 - \sdt)}{\mu
          m}\Big)\right. \nn \\
& +& \left.  \,     \li\Big(1+\frac{(m^2-\pt^2) (m^2-\sdt)}{m^2 \pd^2}\Big) 
  +2 \, \li\Big(1-\frac{\sud}{\pd^2}\Big) +\frac{\pi^2}{12}
+        \ln^2\Big(\frac{\sud}{\pd^2} \frac{(m^2 - \sdt)}{\mu m}\Big) \right\}
+\cO(\e)\,.\nn \\
\eeqa

\subsubsection{Box 10: $I^{D}_4(0,p_2^2,p_3^2,p_4^2;s_{12},s_{23};0,0,0,m^2)$}
\beqa
&&I^{D}_4(0,p_2^2,p_3^2,p_4^2;s_{12},s_{23};0,0,0,m^2)
=\frac{1}{(\sud \sdt-m^2 \sud -\pd^2 \pq^2+m^2 \pd^2)} \nn \\
&\times& \left\{\frac{1}{\e} 
\ln \Big( \frac{(m^2-\pq^2)\pd^2}{(m^2-\sdt)\sud}\Big)
       +\li\Big(1+\frac{(m^2-\pt^2)(m^2-\sdt)}{\pd^2 m^2}\Big)
       -\li\Big(1+\frac{(m^2-\pt^2)(m^2-\pq^2)}{\sud m^2}\Big)
       \right. \nn \\
&+&2 \, \li\Big(1-\frac{m^2-\sdt}{m^2-\pq^2}\Big)
-2 \, \li\Big(1-\frac{\pd^2}{\sud}\Big) 
+2 \, \li\Big(1-\frac{\pd^2(m^2-\pq^2)}{\sud (m^2-\sdt)}\Big) \nn \\
&+&2 \left. \ln \Big( \frac{\mu m}{m^2-\sdt}\Big)
\ln \Big( \frac{(m^2-\pq^2)\pd^2}{(m^2-\sdt)\sud}\Big)
\right\}+\cO(\e)\,.
\eeqa

\subsubsection{Box 11: $I^{D}_4(0,m_3^2,\pt^2,m_4^2;s_{12},s_{23};0,0,m_3^2,m_4^2) $}
\beqa
&&I^{D}_4(0,m_3^2,\pt^2,m_4^2;s_{12},s_{23};0,0,m_3^2,m_4^2) 
=\frac{1}{(m_3^2-\sud)(m_4^2-\sdt)} \nn \\
&\times & \left\{\frac{1}{\e^2} 
-\frac{1}{\e} \ln\bigg(\frac{(m_4^2-\sdt) (m_3^2-\sud) }{m_3 m_4 \mu^2}\Bigg) 
 +2 \ln \Big(\frac{m_3^2-\sud}{m_3 \mu }\Big) \ln
 \Big(\frac{m_4^2-\sdt}{m_4 \mu}\Big)\right.   \nn \\
  &-&\left. \frac{\pi^2}{2}+\ln^2\Big(\frac{m_3}{m_4}\Big)
-\frac{1}{2} \ln^2\Big(\frac{\gp_{34}}{\gp_{34}-1}\Big)
  -\frac{1}{2} \ln^2\Big(\frac{\gm_{34}}{\gm_{34}-1}\Big)
\right\}+\cO(\e)\,,
\eeqa
where 
\beq
\label{eq:gammaij}
\gamma^{\pm}_{ij} = \frac{1}{2}\Bigg[ 1- \frac{m_i^2-m_j^2}{p_3^2} 
\pm \sqrt{(1- \frac{m_i^2-m_j^2}{p_3^2})^2-\frac{4 m_j^2}{p_3^2}}\Bigg]\,.
\eeq
and $\gamma^{+}_{ij}+\gamma^-_{ji}=1$.
Assuming $m_4^2 > m_3^2$, in the limit $\pt^2 \to 0$ we obtain,
\beq
\frac{\gp_{34}}{\gp_{34}-1} \to 1+\cO(\pt^2),\;\;\;
\frac{\gm_{34}}{\gm_{34}-1} \to \frac{m_4^2}{m_3^2}+\cO(\pt^2)\,,
\eeq
and this expression reduces to the form given in eq.~(6.77) of 
H\"opker~\cite{Hopker:1996sx}
\beqa
&&I^{D}_4(0,m_3^2,0,m_4^2;s_{12},s_{23};0,0,m_3^2,m_4^2) 
=\frac{1}{(m_3^2-\sud)(m_4^2-\sdt)} \nn \\
&\times & \left\{\frac{1}{\e^2} 
-\frac{1}{\e} \ln\bigg(\frac{(m_4^2-\sdt) (m_3^2-\sud) }{m_3 m_4
\mu^2}\Bigg)\right.  \nn \\
 &+&\left. 2 \ln \Big(\frac{m_3^2-\sud}{m_3 \mu }\Big) \ln \Big(\frac{m_4^2-\sdt}{m_4 \mu}\Big) 
  -\frac{\pi^2}{2}-\ln^2\Big(\frac{m_3}{m_4}\Big)\right\}+\cO(\e)\,.
\eeqa

\subsubsection{Box 12: $I^{D}_4(0,m_3^2,\pt^2,\pq^2;s_{12},s_{23};0,0,m_3^2,m_4^2) $}
\beqa
&&I^{D}_4(0,m_3^2,\pt^2,\pq^2;s_{12},s_{23};0,0,m_3^2,m_4^2) 
=\frac{1}{(\sud -m_3^2)(\sdt-m_4^2)} \nn \\
&\times & \left\{\frac{1}{2 \e^2} -\frac{1}{\e} \ln\bigg(\frac{(m_4^2-\sdt)(m_3^2-\sud) }{(m_4^2-\pq^2)m_3 \mu}\bigg)
+ 2 \ln\bigg(\frac{m_4^2-\sdt}{m_3 \mu} \bigg)\ln\bigg(\frac{m_3^2-\sud} {m_3 \mu}\bigg)\right. \nn \\
& -&\ln^2\bigg(\frac{m_4^2-\pq^2}{m_3 \mu}\bigg)
 -\frac{\pi^2}{12}
+ 
\ln\Big(\frac{m_4^2-\pq^2}{m_3^2-\sud}\Big)\ln\Big(\frac{m_4^2}{m_3^2}\Big)
-\frac{1}{2} \ln^2\Big(\frac{\gp_{34}}{\gp_{34}-1}\Big)
  -\frac{1}{2} \ln^2\Big(\frac{\gm_{34}}{\gm_{34}-1}\Big)\nn \\
 &-&\left. 2 \,\li\Big(1-\frac{(m_4^2-\pq^2)}{(m_4^2-\sdt)}\Big)
 - \,\li\Big(1-\frac{(m_4^2-\pq^2)}{(m_3^2-\sud)}\frac{\gamma^{+}_{43}}{\gamma^{+}_{43}-1}\Big)
 - \,\li\Big(1-\frac{(m_4^2-\pq^2)}{(m_3^2-\sud)}\frac{\gamma^{-}_{43}}{\gamma^{-}_{43}-1}\Big)
\right\}\nn \\
&+&\cO(\epsilon)\,,
\eeqa
where $\gamma^{\pm}_{ij}$ is given in eq.~(\ref{eq:gammaij}).  
In the limit $\pt^2=0$, we get
\beqa
&&I^{D}_4(0,m_3^2,0,\pq^2;s_{12},s_{23};0,0,m_3^2,m_4^2) 
=\frac{1}{(\sud -m_3^2)(\sdt-m_4^2)} \nn \\
&\times & \left\{\frac{1}{2 \e^2} -\frac{1}{\e} \ln\bigg(\frac{(m_4^2-\sdt)(m_3^2-\sud) }{(m_4^2-\pq^2)m_3 \mu}\bigg)
+ 2 \ln\bigg(\frac{m_4^2-\sdt}{m_3 \mu}
  \bigg)\ln\bigg(\frac{m_3^2-\sud} {m_3 \mu}\bigg)
\right.
\nn \\
& -&\ln^2\bigg(\frac{m_4^2-\pq^2}{m_3 \mu}\bigg)
 -\frac{\pi^2}{12}
+ \ln\Big(\frac{m_4^2-\pq^2}{m_3^2-\sud}\Big)\ln\Big(\frac{m_4^2}{m_3^2}\Big)
-\frac{1}{2}\ln^2\Big(\frac{m_4^2}{m_3^2}\Big)\nn \\
 &-&2 \left. \,\li\Big(1-\frac{(m_4^2-\pq^2)}{(m_4^2-\sdt)}\Big)
 -\,\li\Big(1-\frac{(m_4^2-\pq^2)}{(m_3^2-\sud)}\Big)
 - \,\li\Big(1-\frac{m_3^2}{m_4^2}\frac{(m_4^2-\pq^2)}{(m_3^2-\sud)}\Big)
\right\}
+\cO(\epsilon)\,. \nn \\
\eeqa

\subsubsection{Box 13: $I^{D}_4(0,\pd^2,\pt^2,\pq^2;s_{12},s_{23};0,0,m_3^2,m_4^2) $}
\beqa
&&I^{D}_4(0,\pd^2,\pt^2,\pq^2;s_{12},s_{23};0,0,m_3^2,m_4^2) =\frac{1}{\Delta}
\left\{\frac{1}{\e} \ln\bigg(\frac{(m_3^2-\pd^2)
(m_4^2-\pq^2)}{(m_3^2-\sud) (m_4^2-\sdt)} \bigg)\right. \nn \\
   &-&2\,\li\Big(1-\frac{(m_3^2-\pd^2)}{(m_3^2-\sud)}\Big)
 -\li\Big(1-\frac{(m_3^2-\pd^2)}{(m_4^2-\sdt)}\frac{\gp_{34}}{\gp_{34}-1}\Big) 
 -\li\Big(1-\frac{(m_3^2-\pd^2)}{(m_4^2-\sdt)}\frac{\gm_{34}}{\gm_{34}-1}\Big) 
\nn \\
   &-&2\,\li\Big(1-\frac{(m_4^2-\pq^2)}{(m_4^2-\sdt)}\Big) 
-\li\Big(1-\frac{(m_4^2-\pq^2)}{(m_3^2-\sud)}\frac{\gp_{43}}{\gp_{43}-1} \Big) 
-\li\Big(1-\frac{(m_4^2-\pq^2)}{(m_3^2-\sud)}\frac{\gm_{43}}{\gm_{43}-1} \Big) 
\nn \\
   &+&2\,\li\Big(1-\frac{(m_3^2-\pd^2)(m_4^2-\pq^2)}{(m_3^2-\sud)(m_4^2-\sdt)}\Big)
   +2 \ln\Big(\frac{m_3^2-\sud}{\mu^2}\Big) \ln\Big(\frac{m_4^2-\sdt}{\mu^2}\Big) \nn \\
   &-& \ln^2\Big(\frac{m_3^2-\pd^2}{\mu^2}\Big) -\ln^2\Big(\frac{m_4^2-\pq^2}{\mu^2}\Big) 
+\ln\Big(\frac{m_3^2-\pd^2}{m_4^2-\sdt}\Big)
\ln\Big(\frac{m_3^2}{\mu^2}\Big) 
+\ln\Big(\frac{m_4^2-\pq^2}{m_3^2-\sud}\Big)\ln\Big(\frac{m_4^2}{\mu^2}\Big) \nn \\
&-&\left. \frac{1}{2} \ln^2\Big(\frac{\gp_{34}}{\gp_{34}-1}\Big)
  -\frac{1}{2} \ln^2\Big(\frac{\gm_{34}}{\gm_{34}-1}\Big)
\right\}+\cO(\e)\,,
\eeqa
where $\gamma^{\pm}_{ij}$ is given in eq.~(\ref{eq:gammaij}) and 
\beqa
\Delta &=&(\sud \sdt-m_3^2 \sdt-m_4^2 \sud-\pd^2 \pq^2+m_3^2 \pq^2 +m_4^2 \pd^2)\nn \\
      &=&  (m_3^2-\sud)(m_4^2-\sdt)-(m_3^2-\pd^2)(m_4^2-\pq^2)\,.
\eeqa
In the limit $\pt^2 \to 0$ this simplifies to
\beqa
&&I^{D}_4(0,\pd^2,0,\pq^2;s_{12},s_{23};0,0,m_3^2,m_4^2) =\frac{1}{\Delta}
\left\{\frac{1}{\e} \ln\bigg(\frac{(m_3^2-\pd^2)
(m_4^2-\pq^2)}{(m_3^2-\sud) (m_4^2-\sdt)} \bigg)\right. \nn \\
   &-&2\,\li\Big(1-\frac{(m_3^2-\pd^2)}{(m_3^2-\sud)}\Big)
 -\li\Big(1-\frac{(m_3^2-\pd^2)}{(m_4^2-\sdt)}\Big) 
 -\li\Big(1-\frac{m_4^2}{m_3^2}\frac{(m_3^2-\pd^2)}{(m_4^2-\sdt)}\Big)
\nn \\
   &-&2\,\li\Big(1-\frac{(m_4^2-\pq^2)}{(m_4^2-\sdt)}\Big) 
-\li\Big(1-\frac{(m_4^2-\pq^2)}{(m_3^2-\sud)}\Big) 
-\li\Big(1-\frac{m_3^2}{m_4^2} \frac{(m_4^2-\pq^2)}{(m_3^2-\sud)}\Big) 
\nn \\
   &+&2\,\li\Big(1-\frac{(m_3^2-\pd^2)(m_4^2-\pq^2)}{(m_3^2-\sud)(m_4^2-\sdt)}\Big)
   +2 \ln\Big(\frac{m_3^2-\sud}{\mu^2}\Big) \ln\Big(\frac{m_4^2-\sdt}{\mu^2}\Big) \nn \\
   &-& \ln^2\Big(\frac{m_3^2-\pd^2}{\mu^2}\Big) -\ln^2\Big(\frac{m_4^2-\pq^2}{\mu^2}\Big) 
+\ln\Big(\frac{m_3^2-\pd^2}{m_4^2-\sdt}\Big)
\ln\Big(\frac{m_3^2}{\mu^2}\Big) 
+\ln\Big(\frac{m_4^2-\pq^2}{m_3^2-\sud}\Big)\ln\Big(\frac{m_4^2}{\mu^2}\Big) \nn \\
&-&\left. \frac{1}{2} \ln^2\Big(\frac{m_4^2}{m_3^2}\Big)
\right\}+\cO(\e)\,. 
\eeqa

\subsubsection{Box 14: $I^{D}_4(m_2^2,m_2^2,m_4^2,m_4^2;s_{12},s_{23};0,m_2^2,0,m_4^2) $}
We can obtain this doubly IR divergent box integral 
from eq.~(2.13) of ref.~\cite{Beenakker:1988jr}, using the simple
replacement rule in eq.~(\ref{eq:rep}). We obtain
\beq
I^{D}_4(m_2^2,m_2^2,m_4^2,m_4^2;\sud,\sdt;0,m_2^2,0,m_4^2)=
\frac{-2}{m_2 m_4 \sud}\frac{x_{23} \ln(x_{23})
}{1-x_{23}^2}\left\{\frac{1}{\epsilon}+\ln\Big(\frac{\mu^2}{-\sud}\Big)\right\}+\cO{(\e)},
%\;\;\sdt-(m_2-m_4)^2\neq 0 \nn \\
\eeq
The variable $x_{23}$ is defined in terms of the function $K$,
eq.~(\ref{eq:Kdef}), such that
\beq
\label{eq:x23}
x_{23}=-K(s_{23}+i \varepsilon,m_2,m_4)\,.
\eeq
In the limit $\sdt-(m_2-m_4)^2\to 0$ we have
\beq
I^{D}_4(m_2^2,m_2^2,m_4^2,m_4^2;\sud,\sdt;0,m_2^2,0,m_4^2)=
\frac{1}{ m_2 m_4 \sud } \left\{\frac{1}{\epsilon} 
+\ln\Big(\frac{\mu^2}{-\sud}\Big)\right\}+\cO{(\e,(1-x_{23})^2)}\,. %\;\;\sdt-(m_2-m_4)^2= 0 
\eeq

\subsubsection{Box 15: $I^{D}_4(m_2^2,\pd^2,\pt^2,m_4^2;s_{12},s_{23};0,m_2^2,0,m_4^2) $}
We can obtain  this IR divergent box integral 
from eq.~(2.11) of ref.~\cite{Beenakker:1988jr}, using the simple
replacement rule eq.~(\ref{eq:rep}). We obtain
\beqa
&&I^{D}_4(m_2^2,\pd^2,\pt^2,m_4^2;\sud,\sdt;0,m_2^2,0,m_4^2)= 
\frac{x_{23}}{m_2 m_4 \sud (1-x_{23}^2)}
\nn \\
&\times &\Bigg\{\ln x_{23} \Bigg[ -\frac{1}{\epsilon}
-\frac{1}{2} \ln x_{23} -\ln \Big(\frac{\mu^2}{m_2 m_4}\Big)
- \ln \Big(\frac{m_2^2-\pd^2}{-\sud}\Big)
- \ln \Big(\frac{m_4^2-\pt^2}{-\sud}\Big)\Bigg] \nn \\
&-&\li(1-x_{23}^2)
+\frac{1}{2} \ln^2 y +\sum_{\rho=\pm 1} \li(1-x_{23} y^\rho)
\Bigg\}+\cO{(\e)},
\label{eq:box15}
\eeqa
where 
\beq
y =\frac{m_2}{m_4} \frac{(m_4^2-\pt^2)}{(m_2^2-\pd^2)}\,,
\eeq
and the variable $x_{23}$ is defined in eq.~(\ref{eq:x23}).

For $m_4^2-p_3^2$ small it is useful for numerical purposes to rewrite
eq.~(\ref{eq:box15}) in the form 
\beqa
&&I^{D}_4(m_2^2,\pd^2,\pt^2,m_4^2;\sud,\sdt;0,m_2^2,0,m_4^2) 
=\frac{x_{23}}{m_2 m_4 \sud (1-x_{23}^2)}
\nn \\
&\times & 
\Bigg\{\ln x_{23} \Bigg[ -\frac{1}{\epsilon}
- \ln x_{23} -\ln \Big(\frac{\mu^2}{m_2^2}\Big)
- 2 \ln \Big(\frac{m_2^2-\pd^2}{-\sud}\Big)\Bigg] \nn \\
&-&\li(1-x_{23}^2) +\li(1-x_{23} y)-\li(1-\frac{y}{x_{23}})
\Bigg\}+\cO{(\e)}, 
\eeqa
and similarly for $m_2^2-p_2^2$ small. 

In the limit $x_{23} \to 1$ (i.e $\sdt=(m_2-m_4)^2$) we obtain
\beqa
&&I^{D}_4(m_2^2,\pd^2,\pt^2,m_4^2;\sud,\sdt;0,m_2^2,0,m_4^2)= 
=\frac{1}{2 m_2 m_4 \sud} \nn \\
&\times &
\Bigg\{\frac{1}{\epsilon}+\ln \Big(\frac{\mu^2}{m_2 m_4}\Big)
+ \ln \Big(\frac{m_2^2-\pd^2}{-\sud}\Big)
+\ln \Big(\frac{m_4^2-\pt^2}{-\sud}\Big)-2 - \frac{1+y}{(1-y)} \ln y
\Bigg\}\nn \\ 
&&+\cO{(\e, (1-x_{23})^2)}\,.
\eeqa

\subsubsection{Box 16: $I_4^{D}(m_2^2,\pd^2,\pt^2,m_4^2;s_{12},s_{23};0,m_2^2,m_3^2,m_4^2)$}
We can calculate this IR divergent box integral 
from eq.~(2.9) of ref.~\cite{Beenakker:1988jr}, 
using the simple replacement rule eq.~(\ref{eq:rep}). We obtain
\beqn
&&I_4^{D}(m_2^2,\pd^2,\pt^2,m_4^2;\sud,\sdt;0,m_2^2,m_3^2,m_4^2)
=\frac{x_{23}}{m_2 m_4 (\sud-m_3^2)(1-x_{23}^2)}\nn \\
&\times & 
\Bigg\{-\frac{\ln (x_{23})}{\epsilon}  -2 \ln (x_{23})
\ln \Big(\frac{m_3 \mu }{m_3^2-\sud}\Big)
+ \ln^2(x_2)+ \ln^2(x_3)- \li(1-x_{23}^2) \nn \\
&+& \li(1-x_{23} x_2 x_3)+ \li(1-\frac{x_{23}}{ x_2 x_3})+ \li(1-\frac{x_{23}
x_2} {x_3})+ \li(1-\frac{x_{23} x_3}{x_2})\Bigg\}+\cO{(\e)},\nn\\
\eeqn
where $x_{23} \equiv -K(s_{23},m_2,m_4)$, $x_2 \equiv -K(p_2^2,m_2,m_3)$ and
$x_3 \equiv -K(p_3^2,m_3,m_4)$.

In the limit $x_{23} \to 1$ (i.e $\sdt=(m_2-m_4)^2$) we obtain
\beqa
&&I_4^{D}(m_2^2,\pd^2,\pt^2,m_4^2;\sud,\sdt;0,m_2^2,m_3^2,m_4^2)
=\frac{1}{ 2 m_2 m_4 (\sud-m_3^2)}\nn \\
&\times & 
\Bigg\{\frac{1}{\epsilon}  +2 \ln \Big(\frac{m_3 \mu }{m_3^2-\sud}\Big)
-\frac{1+x_2 x_3}{1-x_2 x_3} \Big[\ln(x_2)+\ln(x_3)\Big]
-\frac{x_3+x_2}{x_3-x_2} \Big[\ln(x_2)-\ln(x_3)\Big] -2 \Bigg\}\nn \\
&+&\cO{(\e, (1-x_{23})^2)}\,.
\eeqn
Special choices of $\pd^2,\pt^2$ and (non-zero) values of the masses
$m_2^2,m_3^2,m_4^2$ will not lead to further divergences.

\begin{table}
\begin{tabular}{|c|l|c|}
\hline
Divergent box &  Special Case & Reference  \\
\hline
Box 8  &  {$I_4^{D}(0,0,p_3^2,p_3^2; s_{12},s_{23};0,0,0,m^2)$} & \cite{Hopker:1996sx}, eq.~(6.71) \\
       & {$I_4^{D}(0,0,0,\pq^2; s_{12},s_{23};0,0,0,m^2)$}      &\cite{Andersen:2007mp}, eq.~(A17) \\
\hline
Box 11 & {$I^{D}_4(0,m_3^2,0,m_4^2;s_{12},s_{23};0,0,m_3^2,m_4^2) $}& \cite{Hopker:1996sx}, eq.~(6.77) \\
       & {$I^{D}_4(0,m^2,0,m^2;s_{12},s_{23};0,0,m^2,m^2) $}& \cite{Hopker:1996sx}, eq.~(6.70) \\
%       & {$I_4^{D}(0,m^2,\pt^2,m^2;s_{12},s_{23},0,0,m^2,m^2)$} &
%       \cite{Rodrigo:1997gv}, eq.~(30),\cite{Beenakker:2002nc}, eq.~(A4), third eqn. \\
       & {$I_4^{D}(0,m^2,\pt^2,m^2;s_{12},s_{23},0,0,m^2,m^2)$} &
       \cite{Rodrigo:1997gv}, eq.~(30) \\
       & &
       \cite{Beenakker:2002nc}, eq.~(A4), third eqn. \\
\hline
Box 12 & {$I^{D}_4(0,m_3^2,0,\pq^2;s_{12},s_{23};0,0,m_3^2,m_4^2)$} &\cite{Berger:2000iu}, eq.~(B7)\\
       & {$I^{D}_4(0,m_3^2,0,m_3^2;s_{12},s_{23};0,0,m_3^2,m_4^2)$} & \cite{Hopker:1996sx}, eq.~(6.74)\\
       & {$I^{D}_4(0,m_3^2,0,\pq^2;s_{12},s_{23};0,0,m_3^2,m_3^2)$} & \cite{Hopker:1996sx}, eq.~(6.78) \\
       & {$\I4^{D}(0,m^2,0,\pq^2;s_{12},s_{23};0,0,m^2,m^2)$} &
       \cite{Beenakker:2002nc}, eq.~(A4), second eqn.\\
       & {$\I4^{D}(0,m^2,\pt^2,\pq^2;s_{12},s_{23},0,0,m^2,m^2)$} &
       \cite{Beenakker:2002nc}, eq.~(A4), fourth eqn.\\
\hline
Box 13 & {$I^{D}_4(0,p^2,0,p^2;s_{12},s_{23};0,0,m^2,m^2)$}  &  \cite{Hopker:1996sx}, eq.~(6.72)  \\
       & {$I^{D}_4(0,m_4^2,0,m_3^2;s_{12},s_{23};0,0,m_3^2,m_4^2)$} & \cite{Hopker:1996sx}  eq.~(6.79) \\
      & $I_4^{D}(0,0,\pt^2,\pq^2; s_{12},s_{23};0,0,m^2,m^2)$
      & \cite{Andersen:2007mp} (v2), eq.~(A19) \\
\hline
Box 16 & {$I_4^{D}(m^2,\pd^2,\pt^2,m^2;s_{12},s_{23};0,m^2,m^2,m^2)$}          & \cite{Beenakker:2002nc}, eq.~(A4), sixth eqn.\\
       & {$I_4^{D}(m^2,0,\pt^2,m^2;s_{12},s_{23};0,m^2,m^2,m^2)$}
       & %\cite{Rodrigo:1997gv},
	\cite{Beenakker:2002nc}, eq.~(A4), fifth eqn.\\
       &  {$I_4^{D}(m^2,0,0,m^2;s_{12},s_{23};0,m^2,m^2,m^2)$}                 & \cite{Beenakker:1988bq}, eq.~(A3) \\
       &  {$I^{D}_4(m_1^2,0,0,m_1^2;s_{12},s_{23};0,m_1^2,m_2^2,m_1^2)$} &\cite{Hopker:1996sx}, eq.~(6.73) \\
       & {$I^{D}_4(m_1^2,0,0,m_2^2;s_{12},s_{23};0,m_1^2,m_1^2,m_2^2)$}  & \cite{Hopker:1996sx}, eq.~(6.76) \\
\hline
\end{tabular}
\caption{Special cases of the 16 basis integrals available in the literature.}
\label{tab:special}
\end{table}

\subsection{Special cases for box integrals}

In this section we give some examples of non-singular limits of the
basis set of box integrals. The first example illustrates that the
basis is overcomplete. If we look at the $m^2\to 0$ limit of box 8,
eq.~(\ref{eq:box8}), we find after little work that it reproduces the
result for box integral 4, eq.~(\ref{eq:box4}). That this should be
the case is clear from the form of the modified Cayley matrix $Y$, as
shown in fig.~\ref{Cayleyfig}. Taking the limit $m^2\to 0$ does not
introduce any new singularities of the form given in
eqs.~(\ref{soft}, \ref{collinear}). In a similar way one can show, for
instance, that box 13 reduces to box 10 in the limit $m_3^2\to0$ and
box 10 goes to box 5 in the limit $m_4^2\to0$.
Since these limits, which are analytically simple, can sometimes be
numerically delicate, we choose to treat these integrals as separate
cases and to categorize the integrals by the number of the internal
masses which are non-vanishing.

The second illustration gives an example of an integral which is
obtainable from one of our basis set integrals by taking a
non-singular limit. If we look at the first entry of
tab.~\ref{tab:special} and take the limit $p_3^2=p_4^2$ we reproduce
the result of ref. \cite{Hopker:1996sx}, eq.~(6.71).  A little care is
required since ref. \cite{Hopker:1996sx} only gives the result for the
real part of the integral in a physical region and has a different
$\epsilon$-dependent overall factor. Table~\ref{tab:special} details
the examples which we have found in the literature which are
non-singular limits of our basis integrals. We have checked that we
are in agreement with all these special cases.

\section{Numerical procedure and checks}

We have constructed a numerical code which for any $N$-point integral
returns the three complex coefficients in the Laurent series
\beq
I_N^D = \frac{a_{-2}}{\e^2}+\frac{a_{-1}}{\e^1}+a_0\,,\hspace{.5cm} N\leq 4 \,. 
\eeq
For the IR divergent triangles and boxes we use the analytic results
of sec.~\ref{sec:results}. For the UV divergent tadpoles and two-point
functions we use the FF library of van Oldenborgh~\cite{van
Oldenborgh:1990yc}. For the finite integrals the coefficients
$a_{-2},a_{-1}$ are equal to zero and we use the FF library for the
coefficient $a_0$. The code for the box integrals is unable to handle
the phase space point where ${\rm det} Y=0$ and one is sitting exactly
at the threshold given by the leading Landau singularity.
The code assumes that all internal masses are real. An extension of
the code which also handles complex masses, appropriate for unstable
particles, is a matter of analytic continuation and programming,
rather than additional calculation.

The code classifies the integrals in terms of the number of non-zero
internal masses and then reduces the integrals to standard forms by
relabelling the integral where necessary. Subsequently on the basis of
the elements of the modified Cayley matrix the code identifies the
appropriate divergent integral and evaluates it, or uses the FF
library.

To perform a numerical check for the divergent box integrals we make
use the identity~\cite{Bern:1992em}
\beq
\label{eq:4to6}
I_4^D = \frac{1}{2} \left( -\sum_{i=1}^4 c_i I_3^D[i]+ (3-D)c_0
I_4^{D+2}\right), 
\eeq
where $I_3^D[i]$ denotes the $D$ dimensional triangle integral
obtained from the box integral $I_4^D$ by removing the $i$-{th}
propagator and the coefficients $c_i$ are given by
\beq
c_{i}= \sum_{j=1}^4 (Y^{-1})_{ij}\,,\hspace{0.5cm}
c_{0}= \sum_{i=1}^4 c_i\,.
\eeq
The six-dimensional box, because it is finite, can be computed
numerically using standard Feynman parameters.  We introduce an
$i\,\varepsilon$ prescription if needed and we set $\varepsilon$ equal
to a small number.
Assuming that the simpler, potentially divergent triangle integrals
have been calculated correctly, using eq.~(\ref{eq:4to6}) we obtain a
rather powerful check of the numerical implementation of the box
integrals, for both space-like and time-like values of the external
invariants $p_i^2$ and $s_{ij}$.

The principal results presented in this paper, as well as the code can
be downloaded from the website
\texttt{http://qcdloop.fnal.gov}.

\section{Conclusions and outlook}
The essential new results of this paper are the classification of the
infrared and collinear divergent triangle and box integrals, the
calculation of the box integrals which were missing from the
literature, and the provision of a code which returns a numerical
answer for any one-loop scalar integral, divergent or finite, for four
or less external legs. We believe that the problem of one-loop scalar
integrals is now completely solved as far as next-to-leading order
calculations are concerned.

In conjunction with a procedure for determining the coefficients with
which scalar integrals with four or less external legs appear in
physical amplitudes we are in principle able to calculate the one-loop
amplitude for any process. Amplitudes with massive internal lines, 
or massless internal lines, or both 
can now be treated in a seamless and uniform way.

\acknowledgments{
We would like to thank Babis Anastasiou, Thomas Becher,
Lance Dixon, Walter Giele, Tim Tait and Stefan Weinzierl for
discussions. RKE thanks the University of Oxford and GZ thanks the ETH
in Z\"urich for hospitality during the writing of this paper.}

\appendix

\section{Useful auxiliary integrals}
In this appendix we report on two integrals which were useful for the
calculation of boxes 11, 12 and
13. The first one is defined as
\beqa
V(\pd^2,\pt^2;\sdt;m_3^2,m_4^2)&=&
\int_0^1 \; d \gamma \;
\frac{(m_3^2-m_4^2-\pd^2+\sdt)}{\gamma(m_3^2-\pd^2)+(1-\gamma)(m_4^2-\sdt)}
\nn\\ 
&&\times \ln\Big(\frac{-\gamma(1-\gamma)\pt^2+\gamma
m_3^2+(1-\gamma)m_4^2}{m_3^2} \Big)\,. 
\eeqa
The result for the integral $V$ is given in terms of the roots $\gamma_{\pm}$ of the quadratic equation
\beq
-\gamma(1-\gamma)\pt^2+\gamma m_3^2+(1-\gamma)m_4^2=0\,. 
\eeq
Hence $\gamma_+ \gamma_-=\frac{m_4^2}{\pt^2}$ and $(1-\gamma_+)
(1-\gamma_-)=\frac{m_3^2}{\pt^2}$ and 
\beq
\gamma_{\pm}=\frac{(\pt^2+m_4^2-m_3^2)\pm \sqrt{(\pt^2+m_4^2-m_3^2)^2-4 m_4^2 \pt^2}}{2 \pt^2}\,.\eeq
Let us define the position of the pole in the integrand as $\gamma_0$,
\beq
\gamma_0=\frac{m_4^2-\sdt}{m_4^2-\sdt-m_3^2+\pd^2}\,. 
\eeq
In terms of these variables $V$ is given by
\beq
V(\pd^2,\pt^2;\sdt;m_3^2,m_4^2)=
\int_0^1 \; d \gamma \; \frac{1}{\gamma-\g0} 
\ln\Big(\frac{(\gamma-\gp)(\gamma-\gm)}{(1-\gp)(1-\gm)}\Big)\,. 
\eeq
With this notation the result for $V$ is
\beqn
\label{eq:Vres}
V(\pd^2,\pt^2;\sdt;m_3^2,m_4^2)&=& 
-\li \Big(1-\frac{\g0-1}{\g0}\frac{\gp}{\gp-1}\Big)
-\li \Big(1-\frac{\g0-1}{\g0}\frac{\gm}{\gm-1}\Big) \nn \\
&+&\li \Big(\frac{1}{1-\gp}\Big)+\li \Big(\frac{1}{1-\gm}\Big)
+2 \;\li \Big(\frac{1}{\g0}\Big)\,. 
\eeqn
If we set $\pd^2=m_3^2$ the pole is at $\g0=1$ and we get
\beq
V(m_3^2,\pt^2;\sdt;m_3^2,m_4^2)= \li \Big(\frac{1}{1-\gp}\Big)+\li
\Big(\frac{1}{1-\gm}\Big)\,. 
\eeq
If we further set $m_3^2=m_4^2=m^2$ we have that $\gp+\gm=1$ and we
obtain  
\beq
V(m^2,\pt^2;\sdt;m^2,m^2)= \li \Big(\frac{1}{1-\gp}\Big)+\li
\Big(\frac{1}{\gp}\Big)= -\frac{1}{2} \ln^2
\Big(\frac{\gp-1}{\gp}\Big)\,. 
\eeq
In the limit $\pt^2 \to 0$ we have 
\beq
\frac{\gp_{34}}{\gp_{34}-1} \to 1+\cO(\pt^2),\;\;\;
\frac{\gm_{34}}{\gm_{34}-1} \to \frac{m_4^2}{m_3^2}+\cO(\pt^2)\,, 
\eeq
and eq.~(\ref{eq:Vres}) simplifies to 
\beqn
V(\pd^2,0;\sdt;m_3^2,m_4^2)&=&
        \li\Big(1-\frac{(m_3^2-\pd^2)}{(m_4^2-\sdt)}\Big)
        +\li\Big(1-\frac{m_4^2}{m_3^2}\Big)\nn \\
        &-&\li\Big(1-\frac{m_4^2 (m_3^2-\pd^2)}{m_3^2 (m_4^2-\sdt)
        }\Big)\,. 
\eeqn
If we further set $\pd^2=m_3^2$ we get
\beq
V(m_3^2,0;\sdt;m_3^2,m_4^2)=\li\Big(1-\frac{m_4^2}{m_3^2}\Big)\,. 
\eeq
Finally, if we also set $m_4^2=m_3^2=m^2$ we obtain 
\beq
V(m^2,0;\sdt;m^2,m^2)=0\,. 
\eeq
Now consider a related integral
\beqa
W(\pd^2,\pt^2;\sdt;m_3^2,m_4^2)&=&
\int_0^1 \; d \gamma \; \frac{(m_3^2-m_4^2-\pd^2+\sdt)}{\gamma(m_3^2-\pd^2)+(1-\gamma)(m_4^2-\sdt)} \nn\\
&&\times \ln\Big(\frac{-\gamma(1-\gamma)\pt^2+\gamma m_3^2+(1-\gamma)m_4^2}{\gamma m_3^2+(1-\gamma)m_4^2} \Big)\,.
\eeqa
Since 
\beq
W(\pd^2,\pt^2;\sdt;m_3^2,m_4^2)=
V(\pd^2,\pt^2;\sdt;m_3^2,m_4^2)-V(\pd^2,0;\sdt;m_3^2,m_4^2)\,,
\eeq
we find
\beqn
W(\pd^2,\pt^2;\sdt;m_3^2,m_4^2)&=&
\li \Big(1-\frac{\g0-1}{\g0}\Big) 
+\li \Big(1-\frac{\g0-1}{\g0}\frac{m_4^2}{m_3^2}\Big)\nn\\
&
-&\li \Big(1-\frac{\g0-1}{\g0}\frac{\gp}{\gp-1}\Big)
-\li \Big(1-\frac{\g0-1}{\g0}\frac{\gm}{\gm-1}\Big) \nn\\
&+&\li \Big(\frac{1}{1-\gp}\Big)+\li \Big(\frac{1}{1-\gm}\Big)-\li\Big(1-\frac{m_4^2}{m_3^2}\Big)\,.
\eeqn

\end{document}